\documentclass[10pt,aps,prd,twocolumn,superscriptaddress,showpacs,nofootinbib]{revtex4-1}
\usepackage{amsmath}
\usepackage{amssymb}
\usepackage{slashed}
\usepackage{graphicx}
\graphicspath{{./figures/}}
\usepackage{float}
\usepackage{physics}
\usepackage{tikz}
\usepackage{tkz-euclide}
\usepackage{siunitx}
\usepackage[colorlinks=true,allcolors=blue]{hyperref}%
\usepackage{xcolor}
\usepackage{comment}
\let\vec\boldsymbol

\newcommand{\Gb}{G} 
\newcommand{\Mb}{M} 
\newcommand{\Rb}{R} 
\newcommand{\Bb}{B} 
\newcommand{\Gf}{\mathcal{G}} 
\newcommand{\Mf}{\mathcal{M}} 
\newcommand{\Rf}{\mathcal{R}} 
\newcommand{\Bf}{\mathcal{B}} 

\tikzset{every picture/.style={line width=0.3mm}}
\definecolor{pred}{RGB}{238,28,37}
\definecolor{pblue}{RGB}{48,49,146}
\definecolor{pgreen}{RGB}{00,163,80}
\usetikzlibrary{decorations.pathmorphing}
\tikzset{snake it/.style={decorate, decoration=snake}}

\begin{document}

\newcommand{\JLU}{Institut f\"ur Theoretische Physik,
  Justus-Liebig-Universit\"at, 
  35392 Giessen, Germany}   
\newcommand{\HFHF}{Helmholtz Research Academy Hesse for FAIR (HFHF), Campus Giessen, 35392 Giessen, Germany}
\newcommand{\UB}{Fakult\"at f\"ur Physik, Universit\"at Bielefeld, D-33615 Bielefeld, Germany}

\title{Effects of dissipation on phase diagram and bosonic excitations\\ 
in the quark-meson model}

\author{Johannes V. Roth}
\affiliation{\JLU}

\author{Yunxin Ye}
\affiliation{\UB}

\author{S\"oren Schlichting}
\affiliation{\UB}

\author{Lorenz von Smekal}
\affiliation{\JLU}
\affiliation{\HFHF}

\date{March 2025}

\begin{abstract}
In this work we study the quark-meson model within a real-time formulation of the functional renormalization group (FRG) on the Schwinger-Keldysh contour. First, we discuss in detail the symmetry of thermal equilibrium for the fermionic sector of the Keldysh action. We take into account dissipation for the bosonic degrees of freedom in the spirit of the Caldeira-Leggett model by coupling the system to an $O(4)$ invariant external heat bath. We study the effect of dissipation on static equilibrium properties, most prominently on the FRG flow of the effective potential and thus on the resulting phase diagram. We find that, unlike in classical systems, through the contributions from non-zero Matsubara modes the dissipative dynamics can in general have an effect on static observables. We investigate these effects within two phenomenological models for the temperature dependence of the pion damping to verify that they are quantitatively small. To estimate their largest possible influence, we consider limits where the damping constants approach infinity. 
\end{abstract}

\maketitle

\section{Introduction}
\label{sec:intro}

One of the long-term goals of ongoing and future heavy-ion collision experiments is an understanding of the QCD phase structure at finite temperature and baryon density. On the theory side, functional methods allow for practical calculations at finite baryon density. These functional methods have offered various predictions on the features of the QCD phase diagram, including the location of the conjectured critical point, the existence of inhomogeneous phases or variants thereof, the equation of the state at large baryon densities, etc.~(see for instance Refs.~\cite{Fischer:2018sdj,Fu:2019hdw,Gao:2020qsj,Gunkel:2021oya,Braun:2022jme}). However, most of these calculations are based on Euclidean field theory. In order to connect these predictions more directly to the phenomenology of heavy-ion collisions, one also needs to understand their impact on the \emph{dynamics} of the rapidly evolving system created in these collisions.

One promising approach to obtain access to the real-time dynamics and spectral properties of strongly interacting systems also from Euclidean field theory is an analytic continuation on the level of the functional equations themselves. This approach avoids the generally ill-posed inverse problem of spectral reconstruction from discrete numerical Euclidean data.
Particularly relevant for the present work are Refs.~\cite{Tripolt:2013jra,Tripolt:2014wra,Jung:2016yxl,Tripolt:2018qvi,Tripolt:2020irx,Tripolt:2021jtp}, where the phase diagram and spectral functions of low-energy effective theories for QCD (including the quark-meson and the parity-doublet model) were computed using the analytically continued functional renormalization group (aFRG). The flow equation of the effective potential was solved in an Euclidean setup, and the phase diagram was computed in the $(\mu,T)$-plane. The spectral functions were computed using analytic continuation, exploiting the 1-loop structure of the flow equations for the two-point functions. For similar approaches based on analytic continuations and/or spectral representations, see e.g.~Refs.~\cite{Pawlowski:2017gxj,Horak:2020eng,Horak:2022aza,Braun:2022mgx,Horak:2022myj,Horak:2023hkp,Horak:2023xfb,Eichmann:2023tjk}.
Despite the merits of such approaches, it is still unclear to which extent one can describe arbitrary real-time dynamics in Euclidean setups. A formulation of the functional renormalization group directly on the Schwinger-Keldysh closed-time path (CTP) \cite{Berges:2012ty,Huelsmann:2020xcy,Roth:2021nrd} provides the possibility of including arbitrary dynamics of the system from the beginning. In the context of dynamic critical phenomena, where one encounters multiple different types of real-time dynamics such as dissipation, diffusion, or reversible mode couplings, this method has been used to study classical-statistical systems such as the relaxational Models A--C \cite{Canet:2006xu,Canet:2011wf,Mesterhazy:2013naa,Mesterhazy:2015uja,Duclut:2016jct,Tan:2021zid,Roth:2023wbp,Chen:2023tqc,Batini:2023nan,Tan:2024fuq}, Model~G \cite{Roth:2024rbi}, and recently also Model~H \cite{Chen:2024lzz,Roth:2024hcu}, in the Halperin-Hohenberg classification \cite{Hohenberg:1977ym}. One of our goals in the present work is to incorporate fermions into the real-time FRG, and use the framework to study the dynamics of the quark-meson model directly from a real-time perspective. A related functional approach is the 2PI formalism, which has been used in Refs.~\cite{Shen:2020jya,Meistrenko:2020nwx} to study   non-equilibrium dynamics in the quark-meson model.

One main difference between the Keldysh formalism and the imaginary-time formalism is that the former allows the system to be out of equilibrium. The special case of the system being in thermal equilibrium instead corresponds to a symmetry of the Keldysh action \cite{Sieberer:2015hba,Altland:2020lbb}, which leads to fluctuation-dissipation relations (FDR's) between real-time correlation functions. Keeping this symmetry intact during the FRG flow constrains the form of the effective average action and therefore simplifies the task of finding suitable truncation schemes. Hence, in the present work, we will generalize the symmetry of thermal equilibrium \cite{Sieberer:2015hba,Altland:2020lbb} to relativistic Dirac fermions, and use it to constrain the fermionic sector of the effective average (Keldysh) action.

Studying the real-time dynamics of the quark-meson model is also worthwhile from the point of view of universal critical dynamics around second-order phase transitions in the QCD phase diagram. While Models~G and H describe at least the universal part of the dynamics of the chiral phase transition in the two-flavor  chiral limit \cite{Rajagopal:1992qz}, and near the QCD critical point \cite{Son:2004iv}, respectively, it is important to also quantify non-universal corrections to the universal scaling behavior. From the point of view of universal dynamics, fermionic excitations are irrelevant, since they always have a Matsubara mass of $\pi T$ and hence never become long-wavelength modes. However, at a large enough `distance' to a second-order phase transition it is important to also include quark dynamics into the effective model picture, in order to more accurately describe the dynamics of QCD.

Having a real-time formulation of the standard quark-meson model at hand, one can start to model the real-time dynamics of the system more accurately. For example, it is known that the chiral order parameter, here described by the $O(4)$ vector $\phi$, is not conserved. To incorporate this into the standard quark-meson model, one can introduce dissipation for $\phi$ in the spirit of the Caldeira-Leggett model \cite{CALDEIRA1983587} by coupling the system to an ensemble of harmonic oscillators and integrating out the latter. This procedure turns the quark-meson model into an open quantum system.
Although this procedure can be performed analogously also in Euclidean spacetime, one encounters the problem that the extraction of real-time quantities such as the associated damping coefficients from the Euclidean effective action, nevertheless requires an analytic continuation again.\footnote{For the damping coefficient one would need the frequency derivative of the retarded self-energy at vanishing external frequency.} In contrast, genuine real-time methods such as the one used in the present work provide direct access to real-time quantities such as the kinetic coefficients and thus are the natural tool for describing dissipative quantum systems.

For classical-statistical systems such as Models A, B, C, etc.~it can be shown within the real-time functional renormalization group method of Ref.~\cite{Roth:2024rbi} that the static sector, described by the free energy, is independent of the dynamic sector, as long as the dynamics drives the system towards a Boltzmann distribution and thus the symmetry of Ref.~\cite{Sieberer:2015hba} applies. As we will see below, the underlying reason is that in a classical-statistical system the only contribution to the static sector comes from the zeroth Matsubara mode, which does not contain any dynamic information of the system. However, for a quantum system such as quark-meson model, this is no-longer the case. 
In this work, we therefore use the quark-meson model as a simple example 
where all Matsubara modes contribute to the flow of the effective action, to demonstrate that the non-zero Matsubara modes generally contain dynamic information of the system. This will then in general also have an effect on the `static sector' described by the effective potential. However, this influence of the dynamics on the static sector is quantitatively small, as we will show comparing the phase diagram and screening masses with zero damping, finite damping and infinite damping.

This paper is organized as follows: In Sec.~\ref{formalism} we introduce the formalism of the real-time quark-meson model using the Schwinger-Keldysh path integral. In particular, the symmetry of thermal equilibrium in a combined system containing both, bosons and fermions is discussed. We then formulate damping terms for sigma and pions which preserve $O(4)$ symmetry, in the spirit of the Caldeira-Leggett model where dissipation is introduced by a coupling to a Gaussian  ensemble of bosonic degrees of freedom~\cite{CALDEIRA1983587}. In Sec.~\ref{sec:floweq}, we derive the real-time FRG flow equation for a system with both fermionic and bosonic fields, and the flow equation for the effective potential of the quark-meson model. The fact that the dynamic information of the system has an influence on the static sector is also discussed. In Sec.~\ref{sct:results}, we present our numerical results on the phase diagram of the quark-meson model as well as the temperature and chemical potential dependence of static observables such as screening masses and dynamic observables such as pole masses without damping, with physically motivated finite damping, and in the limit of infinitely strong over-damping. We show that despite the dynamic information having an influence on the static sector, in general, this influence is quantitatively rather small. In particular, all qualitative features of the phase diagram remain unchanged. In Sec.~\ref{sec:conclusion}, we summarize our findings,  discuss the  conclusions from our work and give an outlook on further studies.

\section{Real-time formalism for the quark-meson model}\label{formalism}
\subsection{Schwinger-Keldysh formalism}\label{sct:QMModelSKContour}
As a low-energy effective model of QCD, the Lagrangian of the quark-meson model with $N_f=2$ flavors of quarks and $N_c=3$ color degrees of freedom is given by
\begin{align}
    \mathcal{L}&= \bar{\Psi}(i\gamma^\mu\partial_\mu -g(\sigma+i\gamma_5 \vec{\tau}\cdot\vec{\pi}))\Psi \label{eq:QMLagr}  \\ \nonumber
    &+\frac{1}{2}((\partial_\mu\sigma)(\partial^\mu\sigma)+(\partial_\mu\vec{\pi})\cdot(\partial^\mu\vec{\pi}))-U_{\Lambda}(\rho)+c\sigma 
\end{align}
where the bare potential $U_{\Lambda}(\rho)$ is defined at the UV scale $\Lambda$ and depends on the $O(4)$ field invariant $\rho = \sigma^2+\vec{\pi}^2$. The explicit symmetry breaking term $-c\sigma$ gives rise to finite (current) quark masses. As usual, we combine $\sigma$ and $\vec{\pi}$ into an $O(4)$ vector $\phi = (\sigma,\vec{\pi})$.

To study the real-time dynamics of the quark-meson model we use the Schwinger-Keldysh closed-time path (CTP) formalism, which we will briefly review in this subsection. The starting point is the Keldysh action, which in continuum notation is given by
\begin{align}
    S = \int_x \left(\mathcal{L}(\bar{\Psi}^+,\Psi^+,\phi^+) - \mathcal{L}(\bar{\Psi}^-,\Psi^-,\phi^-) \right)\,,\label{eq:SKAction}
\end{align}
where the `$+$' and `$-$' fields live on the forward and backward branches of the CTP, respectively. The associated partition function $Z$ is given by \cite{kamenev_2011}
\begin{equation}
    Z = \int \mathcal{D}\bar{\Psi}^+ \mathcal{D}\Psi^+   \mathcal{D}\phi^+ \mathcal{D}\bar{\Psi}^- \mathcal{D}\Psi^- \mathcal{D}\phi^- \,e^{iS} = 1 \,. \label{eq:partFnc}
\end{equation}
For the bosonic fields $\phi$ one can define classical and quantum components with a rotation in Keldysh space,
\begin{align*}
    \phi^c&=\frac{1}{\sqrt{2}}(\phi^++\phi^-) \,, \hspace{0.5cm}
    \phi^q=\frac{1}{\sqrt{2}}(\phi^+-\phi^-) \,.
\end{align*}
For the fermionic fields $\Psi$, it is customary to perform the Keldysh rotation according to\footnote{Note that we do not use the labels `classical' and `quantum' here, but instead simply `1' and `2'. This is because in the Keldysh formalism, fermions (unlike bosons) never have a `classical' meaning in the sense of satisfying a classical equation of motion in the limit $\hbar \to 0$ \cite{kamenev_2011}.}
\begin{align*}
    \Psi_1&=\frac{1}{\sqrt{2}}(\Psi^++\Psi^-) \,, \hspace{0.5cm}
    \Psi_2=\frac{1}{\sqrt{2}}(\Psi^+-\Psi^-) \,.
\end{align*}
In the Larkin-Ovchinnikov convention \cite{kamenev_2011} it is agreed that the conjugate fields $\bar{\Psi}$ transform in the opposite way,\footnote{The conjugate fields $\bar{\Psi}$ are as (Grassmann) variables independent from $\Psi$, so (unlike in the bosonic case) one can choose a different transformation behaviour than for $\Psi$ under the Keldysh rotation.} namely
\begin{align*}
    \bar{\Psi}_1=\frac{1}{\sqrt{2}}(\bar{\Psi}^+-\bar{\Psi}^-) \,, \hspace{0.5cm}
    \bar{\Psi}_2=\frac{1}{\sqrt{2}}(\bar{\Psi}^++\bar{\Psi}^-) \,,
\end{align*}
which has the benefit that the causal structure of the propagator (i.e.~the placement of retarded, advanced, and Keldysh components in the $2\times 2$ matrix, see Eqs.~\eqref{eq:GfInv} and \eqref{eq:Gf} below) is the exactly the same as for its inverse, unlike the bosonic case.

Substituting the Keldysh rotation into \eqref{eq:SKAction} and massaging the terms, the Keldysh action can be written as
\begin{widetext}
\begingroup 
\setlength\arraycolsep{3pt} 
\begin{align}
    S &=  \int_x \bigg[ (\bar{\Psi}_1, \bar{\Psi}_2)
    \begin{pmatrix}
        i\gamma^\mu\partial_\mu -\frac{g}{\sqrt{2}}(\sigma^c+i\gamma_5 \vec{\tau}\cdot\vec{\pi}^c) & -\frac{g}{\sqrt{2}}(\sigma^q+i\gamma_5 \vec{\tau}\cdot\vec{\pi}^q) \\
     -\frac{g}{\sqrt{2}}(\sigma^q+i\gamma_5 \vec{\tau}\cdot\vec{\pi}^q ) & i\gamma^\mu\partial_\mu -\frac{g}{\sqrt{2}}(\sigma^c+i\gamma_5 \vec{\tau}\cdot\vec{\pi}^c)
    \end{pmatrix}
    \begin{pmatrix}
        \Psi_1 \\ \Psi_2
    \end{pmatrix} \label{eq:SKActionQMModel} \\ \nonumber 
    &+\frac{1}{2}(\sigma^c, \sigma^q)
    \begin{pmatrix}
        0 & -\partial^2 \\
    -\partial^2& 0
    \end{pmatrix}
    \begin{pmatrix}
        \sigma^c\\ \sigma^q
    \end{pmatrix}+\frac{1}{2}(\vec{\pi}^c, \vec{\pi}^q )
    \begin{pmatrix}
        0 & -\partial^2 \\
    -\partial^2& 0
    \end{pmatrix}
    \begin{pmatrix}
        \vec{\pi}^c \\ \vec{\pi}^q 
    \end{pmatrix}
    -U_{\Lambda}(\rho^+) + U_{\Lambda}(\rho^-) + \sqrt{2}\,c\sigma^c \bigg] \,,
\end{align}
\endgroup
\end{widetext}
with $\rho^{\pm} \equiv \phi_a^{\pm} \phi_a^{\pm}$ and $\partial^2 \equiv \partial_{\mu}\partial^{\mu}$.
We first discuss the bosonic part of the action. According to the rules of Gaussian integration, the matrix in the quadratic part at some (purely classical, i.e.~$\phi^q = 0$) field expectation value $\phi^c = \sqrt{2}\,\phi$ can be identified with the inverse of the free propagator,
\begin{align}
    \Gb^{-1} &=
    \begin{pmatrix}
        0 & (\Gb^A)^{-1} \\
        (\Gb^R)^{-1}& (\Gb^K)^{-1}
    \end{pmatrix} \nonumber \\
    &= 
    \begin{pmatrix}
        0 & \partial^2+\Mb^2 \\
    \partial^2+\Mb^2& 0
    \end{pmatrix} \label{eq:GbInv}
\end{align}
where we have added a (conventional) additional global minus sign on the right-hand side, and with the mass matrix
\begin{align}
    \Mb^2_{ab} = 2U_{\Lambda}'(\rho) \delta_{ab} + 4\phi_a \phi_b U_{\Lambda}''(\rho) \,.
\end{align}
By inverting \eqref{eq:GbInv} one can then (in principle) find the free propagator,
\begin{align}
    \Gb=\begin{pmatrix}
        \Gb^K & \Gb^R \\
        \Gb^A& 0
    \end{pmatrix} \label{eq:Gb}
\end{align}
However, as it stands, \eqref{eq:GbInv} has no unique inverse. To make its inverse unique, we have to add infinitesimal $\epsilon$-terms which implement the correct causal (retarded and advanced) boundary conditions. In practice, one chooses these infinitesimal $\epsilon$-terms to reproduce the known free retarded/advanced Green functions (see e.g.~\cite{Das:1997gg})
\begin{align}
    \Gb^{R/A}(\omega,\vec{p}) &= -\frac{1}{(\omega\pm i\epsilon)^2 - \vec{p}^2 - M^2} \,, \label{eq:GbRA}
\end{align}
(where the inverse is taken in $4\times 4$ field space).
Moreover, in thermal equilibrium, the Keldysh propagator is generally set by the fluctuation-dissipation relation (FDR)
\begin{align}
    \Gb^K(\omega, \vec{p})=\coth\left(\frac{\omega}{2T}\right)\left(\Gb^R(\omega, \vec{p})-\Gb^A(\omega, \vec{p}) \right) \,. \label{eq:GbKFromFDR}
\end{align} 
Eqs.~\eqref{eq:GbRA} and \eqref{eq:GbKFromFDR} tell us which infinitesimal $\epsilon$-terms we have to add to the action in order for them to be reproduced by the rules of Gaussian integration, namely
\begingroup 
\setlength\arraycolsep{3pt} 
\begin{align}
    &S_b = \label{eq:keldyshActionWithEps} \\  \nonumber
    &\int_x \! \bigg[ \frac{1}{2}(\sigma^c, \sigma^q) \!
    \begin{pmatrix}
        0 & -\partial^2 \!+\!\epsilon u^\mu\partial_\mu\\
    -\partial^2\!-\!\epsilon u^\mu\partial_\mu& -2\epsilon u^\mu\partial_\mu \coth\frac{iu^\mu\partial_\mu}{2T}
    \end{pmatrix} \!
    \begin{pmatrix}
        \sigma^c\\ \sigma^q
    \end{pmatrix} \\ \nonumber
    &+ \!\frac{1}{2}(\vec{\pi}^c, \vec{\pi}^q) \!
    \begin{pmatrix}
        0 & -\partial^2 \!+\! \epsilon u^\mu\partial_\mu \\
    -\partial^2 \!-\! \epsilon u^\mu\partial_\mu& -2\epsilon u^\mu\partial_\mu \coth\frac{iu^\mu\partial_\mu}{2T}
    \end{pmatrix} \!
    \begin{pmatrix}
        \vec{\pi}^c\\ \vec{\pi}^q
    \end{pmatrix}\! \bigg] 
\end{align}
\endgroup
where we have introduced a 4-vector $(u^\mu)=(1,0,0,0)$ to make the action Lorentz covariant. These infinitesimal $\epsilon$-terms ensure that the free bosonic Green functions \eqref{eq:GbRA} and \eqref{eq:GbKFromFDR} are reproduced by the rules of Gaussian integration. In Sec.~\ref{Truncation} it will be shown that the $\epsilon$-coefficient here can be understood as a damping coefficient, i.e.~as a coupling of the system to an external heat bath. At this point, one can imagine that the heat bath is coupled infinitesimally to the system. In this sense, the 4-vector $u^\mu$ can be understood as the 4-velocity of the heat bath.

For the fermionic part of the Schwinger-Keldysh action, because of the different convention of the Keldysh rotation for the conjugate fields $\bar{\Psi}$, the fermionic Green function has a different layout in Keldysh space than the bosonic Green function. The inverse fermionic propagator has the structure (again at constant classical (bosonic) field expectation value $\phi^c = (\sigma^c,\vec{\pi}^c)$, but vanishing quantum field $\phi^q = 0$)
\begin{align}
    \Gf^{-1}&=
    \begin{pmatrix}
        (\Gf^R)^{-1}& (\Gf^K)^{-1} \\
     0 & (\Gf^A)^{-1}
    \end{pmatrix} \nonumber \\
    &= 
    \begin{pmatrix}
        -i\gamma^\mu\partial_\mu +\Mf & 0 \\
     0 & -i\gamma^\mu\partial_\mu +\Mf
    \end{pmatrix}
    \,,  \label{eq:GfInv} 
\end{align}
with $\Mf = \frac{g}{\sqrt{2}}(\sigma^c+i\gamma_5 \vec{\tau}\cdot\vec{\pi}^c)$. The inverse of \eqref{eq:GfInv} (i.e., the free propagator) is given by
\begin{align}
    \Gf=
    \begin{pmatrix}
        \Gf^R& \Gf^K \\
        0 & \Gf^A
    \end{pmatrix} \,. \label{eq:Gf}
\end{align}
Notice that the retarded, advanced, and Keldysh components have the same places as in \eqref{eq:GfInv} due to the the Larkin-Ovchinnikov convention for the Keldysh rotation. It is known that the free fermionic retarded and advanced Green functions should be given by (see e.g.~\cite{Das:1997gg})
\begin{align}
    \Gf^{R/A}(\omega,\vec{p}) &= -\left(\gamma^{0} (\omega\pm i\epsilon) - \vec{\gamma} \cdot\vec{p} - \Mf \right)^{-1} \,. \label{eq:GfRA}
\end{align} 
Moreover, in thermal equilibrium at temperature $T$ and chemical potential $\mu$, the fermionic Keldysh propagator is also generally set by the (fermionic) FDR,
\begin{align}
    \Gf^K(\omega, \vec{p})=\tanh\left(\frac{\omega-\mu}{2T}\right)\left(\Gf^R(\omega, \vec{p})-\Gf^A(\omega, \vec{p}) \right) \,. \label{eq:GfK}
\end{align} 
Eqs.~\eqref{eq:GfRA} and \eqref{eq:GfK} tell us the $\epsilon$-terms which we have to add to action in order for these Green functions to be reproduced by Gaussian integration.

Putting all together, the Keldysh action for the quark-meson model in thermal equilibrium at temperature $T$ and quark chemical potential $\mu$ is given by
\begingroup 
\setlength\arraycolsep{3pt} 
\begin{align}
    S&=  \int_x \bigg[ \\ \nonumber
    &\hspace{-0.7cm} (\bar{\Psi}_1, \bar{\Psi}_2)
    \begin{pmatrix}
        i\gamma^\mu\partial_\mu \!+\! i\epsilon u_\mu \gamma^\mu&  2i\epsilon u_\mu \gamma^\mu\tanh\frac{iu^\mu\partial_\mu-\mu}{2T}\\
     0 & i\gamma^\mu\partial_\mu \!-\! i\epsilon u_\mu \gamma^\mu
    \end{pmatrix}
    \begin{pmatrix}
        \Psi_1 \\ \Psi_2
    \end{pmatrix} \\ \nonumber
     -\frac{g}{\sqrt{2}}
    &(\bar{\Psi}_1,\bar{\Psi}_2)
    \begin{pmatrix}
        \sigma^c \!+\! i\gamma_5 \vec{\tau}\cdot\vec{\pi}^c & \sigma^q \!+\! i\gamma_5 \vec{\tau}\cdot\vec{\pi}^q \\
        \sigma^q \!+\! i\gamma_5 \vec{\tau}\cdot\vec{\pi}^q & \sigma^c \!+\! i\gamma_5 \vec{\tau}\cdot\vec{\pi}^c
    \end{pmatrix}
    \begin{pmatrix}
        \Psi_1 \\ \Psi_2
    \end{pmatrix}\\ \nonumber
    + \frac{1}{2} &(\sigma^c, \sigma^q)
    \begin{pmatrix}
        0 & -\partial^2 \!+\! \epsilon u^\mu\partial_\mu\\
    -\partial^2 \!-\! \epsilon u^\mu\partial_\mu& -2\epsilon u^\mu\partial_\mu \coth\frac{iu^\mu\partial_\mu}{2T}
    \end{pmatrix}
    \begin{pmatrix}
        \sigma^c\\ \sigma^q
    \end{pmatrix} \\ \nonumber 
    +\frac{1}{2} &(\vec{\pi}^c, \vec{\pi}^q)
    \begin{pmatrix}
        0 & -\partial^2 \!+\! \epsilon u^\mu\partial_\mu \\
    -\partial^2 \!-\! \epsilon u^\mu\partial_\mu& -2\epsilon u^\mu\partial_\mu \coth\frac{iu^\mu\partial_\mu}{2T}
    \end{pmatrix}
    \begin{pmatrix}
        \vec{\pi}^c\\ \vec{\pi}^q
    \end{pmatrix} \\ \nonumber
    -&U_{\Lambda}(\rho^+) + U_{\Lambda}(\rho^-) + \sqrt{2}\,c\sigma^c \bigg] \,.
\end{align}
\endgroup

\subsection{The symmetry of thermal equilibrium}
\label{sct:symmThEq}

Before writing down a truncation for the effective average action it is instructive to first look at the symmetries of the bare action. Those symmetries of the bare action that are not violated by source terms or FRG regulators are also symmetries of the effective average action \cite{Gies:2006wv} and can therefore restrict the form of possible truncation schemes.

In the Keldysh formalism systems do not need to be in thermal equilibrium, but if they are, this is reflected in a symmetry of the Keldysh action \cite{Sieberer:2015hba,Altland:2020lbb}. In general, thermal equilibrium implies that expectation values of various numbers of operators, evaluated at arbitrary space-time points, satisfy fluctuation-dissipation relations (FDR's). These FDR's can be shown to be equivalent to the invariance of the Keldysh action under a combination of quantum-mechanical time-reversal (which guarantees the micro-reversibility of the Hamiltonian) and the Kubo-Martin-Schwinger (KMS) conditions (which guarantee that the density operator describes a grand-canonical ensemble) \cite{Sieberer:2015hba,Altland:2020lbb}. 

Before going into the fermionic thermal-equilibrium symmetry, let us first review the argument behind the bosonic thermal-equilibrium symmetry \cite{Sieberer:2015hba}. For a real bosonic wave function $\varphi$,\footnote{Note that the transformations in this subsection are formulated for the (either real or Grassmann valued) fields from the coherent-state path integral, \emph{not} for the field \emph{operators}. This difference will become more apparent in the fermionic case below.} which could represent either $\sigma$ or $\vec{\pi}$, the time-reversal transformation is given by
\begin{align}
    T\varphi(t,\vec{x})=\eta_{\varphi} \, \varphi(-t,\vec{x}) \,,
\end{align}
where $\eta_{\varphi}=\pm 1$ denotes the time parity of the field $\varphi$, i.e.~$\eta_{\varphi}=+1$ for the scalar sigma and $\eta_{\varphi}=-1$ for the pseudoscalar pions. On the CTP, the `$+$' and `$-$' parts of the contour are switched upon time reversal, and thus the time-reversal transformation for bosons on the CTP is given by
\begin{align}
    T\varphi_{\pm}(t,\vec{x})=\eta_{\varphi} \,\varphi_{\mp}(-t,\vec{x})\,.
\end{align}
To formulate the KMS condition on the CTP, one defines a transformation $\mathcal{K}$  according to~\cite{Sieberer:2015hba}
\begin{align}
    \mathcal{K}\varphi_{\pm}(t,\vec{x})= \varphi_{\mp}(t\mp i\beta/2,\vec{x}) \,, \label{eq:defOfK}
\end{align}
with $\beta \equiv 1/T$. However, the transformation in the KMS condition also causes an operator reordering relative to the thermal density operator in correlation functions. To effectively `undo' this relative reordering of operators, one applies another time-reversal transformation. Thermal equilibrium is then expressed as an invariance of the Keldysh action under such a combination $\mathcal{T}=T \mathcal{K}$ of the time-reversal transformation $T$ and the KMS transformation $\mathcal{K}$, i.e.~under
\begin{align}
    \mathcal{T} \varphi_{\pm}(t,\vec{x}) &=  \eta_{\varphi} \,\varphi_{\pm}(-t\pm i\beta/2,\vec{x}) \,.
\end{align}
One can perform the Keldysh rotation and Fourier transformation to frequency/momentum space, and obtain the symmetry of thermal equilibrium which, for our real bosonic fields  $\phi = (\sigma,\vec{\pi})$, is then given by
\begin{align}
    \mathcal{T} \begin{pmatrix}
		\phi^c(\omega,\vec{p}) \\
		\phi^q(\omega,\vec{p})
	\end{pmatrix} &= \\ \nonumber  \label{eq:piThEqSymm}
	&\hspace{-0.8cm} \eta_\varphi \begin{pmatrix}
		\cosh\left(\frac{\omega}{2T}\right) & -\sinh\left(\frac{\omega}{2T}\right) \\
		-\sinh\left(\frac{\omega}{2T}\right) & \cosh\left(\frac{\omega}{2T}\right)
	\end{pmatrix}
	\begin{pmatrix}
		\phi^c(-\omega,\vec{p}) \\
		\phi^q(-\omega,\vec{p})
	\end{pmatrix} \,.
\end{align}

We now turn to the task of generalizing this symmetry to Dirac fermions. First, we need the time-reversal transformation $T$. Time reversal for a Dirac-fermion wave function is given by\footnote{Note again that the transformations here are formulated for the Grassmann-valued fields $\Psi$ in the coherent-state path integral, not for the field operators. Hence, complex conjugation does not imply charge conjugation here. See Chapter 15.13 and especially Eq.~(15.144) in the textbook by Bjorken and Drell~\cite{bjorken1965relativistic} for a discussion about this difference between field operators and wave functions.} \cite{schwabl2008advanced}
\begin{align}
    T\Psi(t,\vec{x})=K\bar{\Psi}^T(-t,\vec{x}) \,,
\end{align}
where the matrix $K$ acts in Dirac space and satisfies 
\begin{equation}
	K^{-1} \gamma^0 K = {\gamma^0}^T \,, \hspace{1.0cm} K^{-1} {\gamma^i} K = -{\gamma^i}^T \,.
\end{equation}
In the  Dirac representation, one e.g.~has $K=\gamma^1 \gamma^3 \gamma^0$.
On the CTP, time reversal exchanges the `$+$' path and `$-$' parts of the contour. For fermionic degrees of freedom one needs to be more careful here, since switching the `$+$' and `$-$' parts on the contour introduces a reverse ordering of Grassmann variables~\cite{Altland:2020lbb}. Taking this into account, the time-reversal transformation for fermions on the CTP is given by 
\begin{align}
    T\Psi_{\pm}(t,\vec{x})=\pm K\bar{\Psi}_{\mp}^T(-t,\vec{x})\,.
\end{align}
The time-reversal transformation for the bar field $\bar{\Psi}$ follows from $T^2=-1$ and reads
\begin{align}
    T\bar{\Psi}_{\pm}(t,\vec{x})=\mp \Psi_{\mp}^T(-t,\vec{x})K^{-1} \,.
\end{align}
Next, we need the transformation $\mathcal{K}$ used in the KMS condition. At zero chemical potential $\mu=0$ the transformation $\mathcal{K}$ is given just as in the bosonic case,
\begin{subequations}
\begin{align}
    \mathcal{K}\Psi_{\pm}(t,\vec{x})&= \Psi_{\mp}(t\mp i\beta/2,\vec{x})\,,\\
    \mathcal{K}\bar{\Psi}_{\pm}(t,\vec{x})&= \bar{\Psi}_{\mp}(t\mp i\beta/2,\vec{x})\,.
\end{align}
\end{subequations}
At non-vanishing chemical potential $\mu \neq 0$, the transformation $\mathcal{K}$ is generalised to
\begin{subequations}
\begin{align}
    \mathcal{K}\Psi_{\pm}(t,\vec{x})&= e^{\pm \beta\mu/2}\Psi_{\mp}(t\mp i\beta/2,\vec{x}) \,, \\
    \mathcal{K}\bar{\Psi}_{\pm}(t,\vec{x})&= e^{\mp \beta\mu/2}\Psi_{\mp}(t\mp i\beta/2,\vec{x}) \,.
\end{align}
\end{subequations}
Just as in the bosonic case, the symmetry of thermal equilibrium is again equivalent to the invariance under the combination $\mathcal{T}=T\mathcal{K}$ of time reversal and KMS transformation \cite{Altland:2020lbb}, and is explicitly given by
\begin{subequations}
\begin{align}
    \mathcal{T}\Psi_{\pm}(t,\vec{x}) &= \pm e^{\pm \beta\mu/2} \,K \bar{\Psi}_{\pm}^T\left(-t \pm i/(2T),\vec{x}\right) \,, \\
    \mathcal{T}\bar\Psi_{\pm}(t,\vec{x}) &= \mp e^{\mp \beta\mu/2} \, \Psi_{\pm}^T(-t \pm i/(2T),\vec{x}) K^{-1} \,. 
\end{align}
\end{subequations}
Finally, one can perform the Keldysh rotation and a Fourier transformation to frequency and momentum space to obtain
\begin{align}
    \mathcal{T} \begin{pmatrix}
        \Psi_1(\omega,\vec{p}) \\
        \Psi_2(\omega,\vec{p})
    \end{pmatrix} &= \label{eq:fermionicThEqSymmFiniteMu1} \\ \nonumber 
    &\hspace{-2.0cm} \begin{pmatrix}
        \cosh\left(\frac{\omega-\mu}{2T}\right) & -\sinh\left(\frac{\omega-\mu}{2T}\right) \\
        -\sinh\left(\frac{\omega-\mu}{2T}\right) & \cosh\left(\frac{\omega-\mu}{2T}\right)
    \end{pmatrix}
    \begin{pmatrix}
        K \bar{\Psi}_1^T(-\omega,\vec{p}) \\
        K \bar{\Psi}_2^T(-\omega,\vec{p})
    \end{pmatrix} \,, \\
    \mathcal{T} \begin{pmatrix}
        \bar{\Psi}_1(\omega,\vec{p})\\
        \bar{\Psi}_2(\omega,\vec{p})
    \end{pmatrix} &=  \label{eq:fermionicThEqSymmFiniteMu2} \\ \nonumber 
    &\hspace{-2.0cm} \begin{pmatrix}
        -\cosh\left(\frac{\omega+\mu}{2T}\right) & \sinh\left(\frac{\omega+\mu}{2T}\right) \\
        \sinh\left(\frac{\omega+\mu}{2T}\right) & -\cosh\left(\frac{\omega+\mu}{2T}\right) 
    \end{pmatrix}
    \begin{pmatrix}
        \Psi_1^T(-\omega,\vec{p}) K^{-1} \\
        \Psi_2^T(-\omega,\vec{p}) K^{-1}
    \end{pmatrix} \,. 
\end{align}

One can explicitly verify that the Keldysh action \eqref{eq:SKActionQMModel} for the quark-meson model indeed satisfies the combined thermal-equilibrium symmetry of bosons and fermions, which means that it is invariant under the combined transformation $\mathcal{T}$.
It can also be shown that the invariance under $\mathcal{T}$ leads to the well-known FDR
\begin{align}
    \Gb^K(\omega, \vec{p})=\coth\left(\frac{\omega}{2T}\right)\left(\Gb^R(\omega, \vec{p})-\Gb^A(\omega, \vec{p})\right) \label{eq:bosonicFDR}
\end{align}
for the bosonic propagator, and
\begin{align}
    \Gf^K(\omega, \vec{p})=\tanh\left(\frac{\omega-\mu}{2T}\right)\left(\Gf^R(\omega, \vec{p})-\Gf^A(\omega, \vec{p})\right) \label{eq:fermionicFDR}
\end{align}
for the fermionic propagator.

\subsection{Bosonic heat bath}
\label{sct:bosonicHeatBath}

To study the dissipative dynamics of the chiral order parameter,  we consider the $O(4)$ vector $\phi$ to be coupled to an external heat bath. A coupling of the mesons to a heat bath will introduce finite decay widths in their respective spectral functions. In the vacuum, it is well known that the decay width of the sigma is much bigger than that of the pion. Our goal in this subsection is therefore to model the heat bath in a way that allows different decay widths for $\sigma$ and $\vec{\pi}$ in the symmetry-broken phase, while simultaneously keeping the $O(4)$ symmetry of the effective average action intact. As a further requirement, the additional terms added to the Keldysh action should satisfy the symmetry of thermal equilibrium from Sec.~\ref{sct:symmThEq} above. 
In order to meet these requirements, we consider a physical picture where the mesonic part of the action is coupled to an ensemble of harmonic oscillators in the spirit of the Caldeira-Leggett model~\cite{CALDEIRA1983587}.

As a starting point, we consider additional interaction terms in the Lagrangian \eqref{eq:QMLagr} which describe the coupling between the mesons and two new sets of heat-bath oscillators $\{\varphi_s\}$ and $\{\chi_{s,a}\}$ (enumerated by $s$), where the former are $O(4)$ scalars and the latter $O(4)$ vectors. On the level of the Lagrangian \eqref{eq:QMLagr} they couple to the system via 
\begin{equation}
    \mathcal{L}_{int} = \sum_s \frac{g_s}{2} \varphi_s \phi_a \phi_a  + \sum_{s} h_{s} \chi_{s,a}\phi_a  \,,
\end{equation}
with two sets of coupling constants $\{g_s\}$ and $\{h_{s}\}$. This extra piece of the interacting Lagrangian will introduce extra terms in the Keldysh action in the following way (after Keldysh rotation),
\begin{align}
    S_{int} 
    &= \int_x \bigg[ \sum_s \frac{g_s}{\sqrt{2}} \big( \varphi_s^c \phi_a^c \phi_a^q + \frac{1}{2}\varphi_s^q \phi_a^c \phi_a^c + \frac{1}{2}\varphi_s^q\phi_a^q \phi_a^q \big) \; + \nonumber \\
    &\hspace{2.0cm} \sum_{s} h_{s}\big( \chi_{s,a}^c \phi^q_a + \chi_{s,a}^q \phi^c_a \big) \bigg] \,.
\end{align}
For later convenience, in addition to the standard field invariant $\rho = \frac{1}{2}\phi^c_a \phi^c_a$, we introduce the shorthand notation $\rho^q \equiv \phi^c_a \phi^q_a$ and $\rho^{qq} \equiv \frac{1}{2}\phi^q_a \phi^q_a$ for the other $O(4)$ field invariants. 
At the leading order in fluctuations $\delta\phi$ around the field expectation value $\phi_0^c = (\sqrt 2 \sigma_0,0)$ these reduce to  $\rho = \sigma_0^2+ \sqrt 2 \sigma_0\, \delta \sigma^c + \mathcal O(\delta\phi^2)$ and  $\rho^q = \sqrt 2\sigma_0\, \delta \sigma^q + \mathcal O(\delta\phi^2)$, whereas $\rho^{qq} \sim \mathcal O(\delta\phi^2)$ yields true quantum corrections which are absent from the classical MSR action.

Then there is also the non-interacting (kinetic) part of the Lagrangian for the heat-bath oscillators, which specifies their real-time Green functions,
\begin{widetext}
\begin{align}
    S_{bath} = -\frac{1}{2}\int_{\omega\vec{p}} \bigg[ 
    &\sum_s \left(\varphi_s^c,\varphi_s^q\right)_{-\omega,-\vec{p}}
    \begin{pmatrix}
        0 & [D_{\varphi,s}^{-1}]^{R}(\omega,\vec{p}) \\
        [D_{\varphi,s}^{-1}]^{A}(\omega,\vec{p}) & [D_{\varphi,s}^{-1}]^{K}(\omega,\vec{p})
    \end{pmatrix}
    \begin{pmatrix}
        \varphi_s^c \\ \varphi_s^q
    \end{pmatrix}_{\omega,\vec{p}} \; \\ \nonumber 
    &\sum_{s} \left(\chi_{s,a}^c,\chi_{s,a}^q\right)_{-\omega,-\vec{p}}
    \begin{pmatrix}
        0 & [D_{\chi,s}^{-1}]^{R}(\omega,\vec{p}) \\
        [D_{\chi,s}^{-1}]^{A}(\omega,\vec{p}) & [D_{\chi,s}^{-1}]^{K}(\omega,\vec{p})
    \end{pmatrix}
    \begin{pmatrix}
        \chi_{s,a}^c \\ \chi_{s,a}^q
    \end{pmatrix}_{\omega,\vec{p}}
    \bigg] \,.
\end{align}
\end{widetext}
We require the heat bath to be in thermal equilibrium, which mean that the real-time Green functions of the heat-bath oscillators are related by the FDR
\begin{subequations}
\begin{align}
    D_{\varphi,s}^K(\omega,\vec{p})&=\coth\left(\frac{\omega}{2T}\right)\left(D_{\varphi,s}^R(\omega,\vec{p})-D_{\varphi,s}^A(\omega,\vec{p})\right)\,, \\
    D_{\chi,s}^K(\omega,\vec{p})&=\coth\left(\frac{\omega}{2T}\right)\left(D_{\chi,s}^R(\omega,\vec{p})-D_{\chi,s}^A(\omega,\vec{p})\right)\,.
\end{align}
\end{subequations}
Since the heat-bath oscillators $\varphi_s$ and $\chi_{s,a}$ are quadratic in the Keldysh action, one can integrate them out and obtain the terms
\begin{align}
    &S \to S 
    +\int_{\omega\vec{p}} \bigg\{\\
    &\sum_{s} \! \frac{h_{s}^2}{2} \left(\phi_a^c,\phi_a^q\right)_{-\omega,-\vec{p}} \!
    \begin{pmatrix}
        0 & D_{\chi,s}^A(\omega,\vec{p}) \\
        D_{\chi,s}^R(\omega,\vec{p}) & D_{\chi,s}^K(\omega,\vec{p}) 
    \end{pmatrix} \!
    \begin{pmatrix}
        \phi_a^c \\
        \phi_a^q
    \end{pmatrix}_{\omega,\vec{p}} \nonumber \\
    &+\sum_s \frac{g_s^2}{4} \big[  \rho^q(-\omega,-\vec{p}) \, D_{\varphi,s}^R(\omega,\vec{p}) \, \left(\rho(\omega,\vec{p}) + \rho^{qq}(\omega,\vec{p}) \right) \nonumber \\
    &\hspace{0.5cm} + \left(\rho(-\omega,-\vec{p})+\rho^{qq}(-\omega,-\vec{p})\right) \, D_{\varphi,s}^A(\omega,\vec{p}) \, \rho^q(\omega,\vec{p}) \nonumber \\[0.15cm]
    &\hspace{2.0cm} +\rho^q(-\omega,-\vec{p}) \, D_{\varphi,s}^K(\omega,\vec{p}) \, \rho^q(\omega,\vec{p}) \big] \bigg\} \,,
\end{align}
where we have reorganized the terms in the first line (coming from the $\chi$-oscillators) into a quadratic form in $\phi^c$ and $\phi^q$ to emphasize the structure of the self energy which the system particle acquires due to mixing with the $\chi$-oscillators. 
One can introduce spectral densities $J_{\varphi}(\omega,\vec{p})$ and $J_{\chi}(\omega,\vec{p})$ to rewrite the retarded and the advanced heat-bath propagators as
\begin{subequations}
\begin{align}
    \sum_s g_s^2 D_{\varphi,s}^{R/A}(\omega,\vec{p})  &= - \int_0^\infty \frac{d\omega'}{2\pi} \frac{2\omega' J_{\varphi}(\omega',\vec{p})}{(\omega\pm i\epsilon)^2-\omega'^2} \,,\\
    \sum_{s} h_{s}^2 D_{\chi,s}^{R/A}(\omega,\vec{p})  &= - \int_0^\infty \frac{d\omega'}{2\pi} \frac{2\omega' J_{\chi}(\omega',\vec{p})}{(\omega\pm i\epsilon)^2-\omega'^2} \,.
\end{align} \label{eq:hbProp}%
\end{subequations}
Using the FDR, the Keldysh components of the heat-bath propagators are given by
\begin{align}
    \sum_s g_s^2 D^K_s(\omega,\vec{p}) &= \sum_s g_s^2 \coth\left(\frac{\omega}{2T}\right)\,2i\,\Im D^R_s(\omega,\vec{p}) \nonumber \\
    &=\coth\left(\frac{\omega}{2T}\right)iJ_{\varphi}(\omega,\vec{p}) \,, \\
    \sum_{s} h_{s}^2 D^K_{s}(\omega,\vec{p}) &= \sum_{s} h_{s}^2 \coth\left(\frac{\omega}{2T}\right)\,2i \Im D^R_{s}(\omega,\vec{p}) \nonumber \\
    &=\coth\left(\frac{\omega}{2T}\right)iJ_{\chi}(\omega,\vec{p}) \,,
\end{align}
In this work, we consider Ohmic spectral densities for both ensembles of heat-bath oscillators,
\begin{align}
    J_{\chi}(\omega,\vec{p})=2\gamma \omega \;\;\;\text{and}\;\;\; J_{\varphi}(\omega,\vec{p})=2 Y \omega\,, \label{eq:ohmicJgammaY}
\end{align}
with two different damping coefficients $\gamma$ and $Y$.
With this ansatz, after absorbing an infinite constant at $\omega=0$, $\vec{p}=\vec{0}$ into renormalized squared mass and quartic coupling, one obtains
\begingroup 
\setlength\arraycolsep{7pt} 
\begin{align}
    &S \to S \;+ \\ \nonumber 
    &\int_{\omega\vec{p}} \bigg\{ \frac{1}{2}
    \left(\phi_a^c,\phi_a^q\right)_{-\omega,-\vec{p}}
    \begin{pmatrix}
       0 & -i\gamma\omega \\
       i\gamma\omega & 2i\gamma\omega \coth\left(\frac{\omega}{2T}\right) 
    \end{pmatrix}
    \begin{pmatrix}
       \phi_a^c \\
       \phi_a^q
    \end{pmatrix}_{\omega,\vec{p}} \\ \nonumber 
    &\hspace{1.0cm} +\frac{1}{2} \rho^q(-\omega,-\vec{p}) \, \,iY\omega \, \big(\rho(\omega,\vec{p}) + \rho^{qq}(\omega,\vec{p}) \big) \\ \nonumber
    &\hspace{1.0cm} +\frac{1}{2} \rho^q(-\omega,-\vec{p}) \, \,iY\omega \coth\left(\frac{\omega}{2T}\right) \, \rho^q(\omega,\vec{p})
     \bigg\} \,,
\end{align}
\endgroup
for the additional terms in the Keldysh action.
Undoing the Fourier transformation, we can therefore write
\begin{widetext}
\begingroup 
\setlength\arraycolsep{7pt} 
\begin{align}
    S \to S 
    + \int_{x} \bigg\{ &\frac{1}{2} \left(\phi_a^c(x),\phi_a^q(x)\right)
    \begin{pmatrix}
        0 & \gamma u^{\mu}\partial_{\mu} \\
        -\gamma u^{\mu}\partial_{\mu} & -2\gamma u^{\mu}\partial_{\mu} \coth\left(\frac{iu^{\mu}\partial_{\mu}}{2T}\right) 
    \end{pmatrix}
    \begin{pmatrix}
        \phi_a^c(x) \\
        \phi_a^q(x)
    \end{pmatrix} \nonumber \\
    -&\frac{1}{2} \rho^q(x) \, Y u^{\mu}\partial_{\mu} \, \big(\rho(x)+\rho^{qq}(x)\big)
    -\frac{1}{2}  \rho^q(x) \, Y u^{\mu}\partial_{\mu} \coth\left(\frac{iu^{\mu}\partial_{\mu}}{2T}\right) \, \rho^q(x)\bigg\} \,, \label{eq:bareActionHB}
\end{align}
\endgroup
\end{widetext}
where we have again introduced the 4-velocity $u^\mu$ of the heat bath, given by $(u^\mu)=(1,0,0,0)$ in the local rest frame, to make the action Lorentz covariant. Eq.~\eqref{eq:bareActionHB} constitutes our model of an Ohmic heat bath  that produces  different damping constants for the longitudinal  and transverse components in field space while keeping the action $O(4)$ invariant. Specifying $\phi_0^c = (\sqrt 2 \sigma_0,0)$ for symmetry breaking in the direction of the sigma meson, from the imaginary parts of the inverses of the retarded propagators of pions and sigma meson at tree level, with 
\begin{align*}
  &  \rho^q  Y  u^\mu \partial_\mu (\rho + \rho^{qq}) = \delta\sigma^q  2\sigma_0^2 \, Y  u^\mu \partial_\mu  \delta\sigma^c + \mathcal O (\delta\phi^3)\, , \;\; \mbox{and}\\
  &  \rho^q  Y  u^\mu \partial_\mu \coth\left(\frac{iu^{\mu}\partial_{\mu}}{2T}\right) \, \rho^q = \\
  &\hskip 1cm  \delta\sigma^q 2 \sigma_0^2\,  Y  u^\mu \partial_\mu \coth\left(\frac{iu^{\mu}\partial_{\mu}}{2T}\right) \, \delta\sigma^q   + \mathcal O (\delta\phi^3)\, , 
\end{align*}
we can identify their respective damping constants as
\begin{align}
    \gamma_{\pi}=\gamma  \;\;\;\text{and}\;\;\; \gamma_{\sigma} = \gamma+  \sigma_0^2\, Y   
    \,. \label{eq:gamPiAndGamSigma}
\end{align}
Thus our model \eqref{eq:bareActionHB} indeed leads to different damping constants for sigma meson and pions, with the difference $\gamma_{\sigma} - \gamma_{\pi} \sim \sigma_0^2 $ being proportional to the squared symmetry-breaking field expectation value.

\section{Real-time functional renormalization group with fermions}
\label{sec:floweq}

The functional renormalization group (FRG) is a powerful tool which allows one to incorporate both bosonic and fermionic fluctuations non-perturbatively. The standard (Euclidean) formulation of the FRG has been applied rather successfully in the past to study the phase structure of the quark-meson model and variants thereof, e.g., see Refs.~\cite{Berges:1997eu,Schaefer:2004en,Strodthoff:2011tz,Tripolt:2013jra,Tripolt:2014wra,Jung:2016yxl,Tripolt:2017zgc,Resch:2017vjs,Tripolt:2018qvi,Tripolt:2020irx,Tripolt:2021jtp,Grossi:2021ksl,Otto:2022jzl,Ihssen:2023xlp}. However, genuine real-time formulations of the FRG on the Schwinger-Keldysh contour have been applied mostly to quantum systems with only bosonic degrees of freedom in the past,  e.g., in Refs.~\cite{Berges:2012ty,Mesterhazy:2015uja,Huelsmann:2020xcy,Tan:2021zid,Roth:2021nrd,Roth:2023wbp}. The purpose of the present section is twofold: we extend the real-time FRG to include fermionic degrees of freedom, and establish some general notation used in this work.

To formulate the FRG flow equation we follow the standard procedure of adding an infrared regulator $\Delta S_k$ to the bare Keldysh action \eqref{eq:SKAction},
\begin{equation}
    S \to S + \Delta S_k \,.\label{eq:addRegulator}
\end{equation}
The regulator term $\Delta S_k$ has the purpose of suppressing the fluctuations of modes with momentum smaller than the FRG scale~$k$. In the simplest case the regulator term $\Delta S_k$ can be taken to be a quadratic form of the fields, i.e.
\begin{align}
    &\Delta S_k = \label{eq:regDef} \\ \nonumber 
    \frac{1}{2}&\int_{xx'} \!\!\!\! \left(\phi_a^c(x), \phi_a^q(x)\right) \!
    \begin{pmatrix}
        \Rb_{ab,k}^{\tilde{K}}(x,x') & \Rb_{ab,k}^A(x,x') \\
        \Rb_{ab,k}^{R}(x,x') & \Rb_{ab,k}^K(x,x')
    \end{pmatrix} \!
    \begin{pmatrix}
        \phi_b^c(x') \\
        \phi_b^q(x')
    \end{pmatrix} \\ \nonumber + 
    &\int_{xx'} \!\!\!\! \left(\bar{\psi}_1(x), \bar{\psi}_2(x)\right) \!
    \begin{pmatrix}
        \Rf_k^R(x,x') & \Rf_k^K(x,x') \\
        \Rf_k^{\widetilde{K}}(x,x') & \Rf_k^A(x,x')
    \end{pmatrix} \!
    \begin{pmatrix}
        \psi_1(x') \\
        \psi_2(x')
    \end{pmatrix} \,. 
\end{align}
To simplify the notation, one can introduce Keldysh indices $i,j$ (with $i,j=1,2$ for fermions and $i,j=c,q$ for bosons) to write the regulator term as
\begin{align}
    \Delta S_k &= \int_{xx'} \bigg[ \frac{1}{2}\phi_{a}^i(x) \Rb_{ab,k}^{ij}(x,x') \phi_{b}^j(x') \;+ \label{eq:delSk} \\ \nonumber
    &\hspace{4.0cm} \bar{\psi}_i(x) \Rf_{ij,k}(x,x') \psi_j(x') \bigg]  \,. 
\end{align}
Adding the regulator term \eqref{eq:addRegulator} to the bare action means that the generating functional \eqref{eq:Z} and the Schwinger functional \eqref{eq:W}, both explicitly given in Appendix \ref{sct:EKA} to introduce our notations and conventions, become dependent on the FRG scale~$k$: $Z \to Z_k$, $W \to W_k$. One then defines the effective \emph{average} action $\Gamma_k$ as a `modified' Legendre transform of the now scale-dependent Schwinger functional $W_k$, via 
\begin{align}
    \Gamma_k &= W_k - \Delta S_k \;- \label{eq:definition_eff_avg_action} \\ \nonumber
    &\hspace{0.5cm} \int_x \big[ j_a^q \phi_a^c + j_a^c \phi_a^q + \bar{\eta}_1\psi_2 + \bar{\psi}_2\eta_1 + \bar{\eta}_2\psi_1 + \bar{\psi}_1\eta_2 \big] \, .
\end{align}
The effective average action $\Gamma_k$ can be interpreted as the effective action \eqref{eq:Gam} but with all modes of momenta larger than the FRG scale $k$ integrated out. The flow of $\Gamma_k$ can then be derived following the standard procedure by Wetterich \cite{Wetterich:1992yh}. The result here reads
\begin{equation}
    \partial_k \Gamma_k =  i \Tr \left\{ \partial_k \Rf_{k}\circ \Gf_{k} \right\} - \frac{i}{2} \Tr \left\{ \partial_k \Rb_{k}\circ \Gb_{k} \right\} \,.\label{eq:floweq}
\end{equation}
The full field-dependent fermionic propagator $\Gf_k$ is related to the inverse two-point function $\Gamma_k^{(2)}$ via
\begin{align}
    \Gf_{ij,k}(x,x') &=   -\left[ \frac{\overrightarrow{\delta}}{\delta \bar{\psi}_i(x)} \Gamma_k \frac{\overleftarrow{\delta}}{\delta \psi_j(x')} + \Rf_{ij,k}(x,x') \right]^{-1}.
\end{align}
where $i,j=1,2$ are the indices in Keldysh space with Larkin-Ovchinnikov conventions again. Similarly, the full field-dependent bosonic propagator $\Gb_k$ is given by the inverse 
\begin{align}
    \Gb_{ab,k}^{ij}(x,x') &=  -\left[ \frac{\delta^2 \Gamma_k}{\delta \phi_{a}^{i}(x) \delta \phi_{b}^{j}(x')}   + \Rb_{ab,k}^{ij}(x,x') \right]^{-1}
\end{align}
with $i,j=c,q$ here denoting the classical/quantum components in Keldysh space, and $a,b=1,\ldots,4$ the $O(4)$ indices of the corresponding field components.

For the rest of this paper, we assume that both the bosonic and fermionic regulator terms in \eqref{eq:delSk} are chosen to comply with the causal structure of the Keldysh action \cite{kamenev_2011}. As discussed in detail in Sec.~2.1 of Ref.~\cite{Roth:2023wbp}, this 
implies that the retarded/advanced components of the regulators admit spectral representations, and that their anomalous components $\Rb_k^{\tilde{K}}=0$ and $\Rf_k^{\tilde{K}}=0$ vanish. Moreover, we assume that the regulators themselves satisfy the symmetries of thermal equilibrium from Sec.~\ref{sct:symmThEq}. This means that they must be spacetime-translation invariant, and that their respective Keldysh components are set by the FDR, i.e.~after Fourier transformation,
\begin{align}
    \Rb_{ab,k}^K(\omega, \vec{p})=\coth\left(\frac{\omega}{2T}\right)\left(\Rb_{ab,k}^R(\omega, \vec{p})-\Rb_{ab,k}^A(\omega, \vec{p})\right) \label{eq:bosonicFDRReg}
\end{align}
and
\begin{align}
    \Rf_k^K(\omega, \vec{p})=\tanh\left(\frac{\omega-\mu}{2T}\right)\left(\Rf_k^R(\omega, \vec{p})-\Rf_k^A(\omega, \vec{p})\right) \,. \label{eq:fermionicFDRReg}
\end{align}
For later use we also define,
\begin{subequations}
\begin{align}
    \Bb_{ab,k}^R (x,&x') \equiv \\ \nonumber
    \int_{yy'} &\Gb_{ac,k}^R(x,y) \,\partial_k \Rb_{cd,k}^R(y,y') \, \Gb_{db,k}^R(y',x') \,, \\
    \Bb_{ab,k}^A(x,&x') \equiv \\ \nonumber
    \int_{yy'} &\Gb_{ac,k}^A(x,y) \,\partial_k \Rb_{cd,k}^A(y,y') \, \Gb_{db,k}^A(y',x') \,, \\
    \Bb_{ab,k}^K (x,&x') \equiv \\ \nonumber
     \int_{yy'} \Big[ & \Gb_{ac,k}^R(x,y) \,\partial_k \Rb_{cd,k}^R(y,y') \, \Gb_{db,k}^K(y',x') \;+ \\ \nonumber
    &\Gb_{ac,k}^K(x,y) \,\partial_k \Rb_{cd,k}^A(y,y') \, \Gb_{db,k}^A(y',x') \;+  \\ \nonumber
    &\Gb_{ac,k}^R(x,y) \,\partial_k \Rb_{cd,k}^K(y,y') \, \Gb_{db,k}^A(y',x') \Big] \,,
\end{align}
\end{subequations}
and, analogously,
\begin{subequations}
\begin{align}
    \Bf_{k}^R(x,x') &\equiv \int_{yy'} \Gf_{k}^R(x,y) \,\partial_k \Rf_{k}^R(y,y') \, \Gf_{k}^R(y',x') \,, \\
    \Bb_{k}^A(x,x') &\equiv \int_{yy'} \Gf_{k}^A(x,y) \,\partial_k \Rf_{k}^A(y,y') \, \Gf_{k}^A(y',x') \,, \\
    \Bf_{k}^K(x,x') &\equiv \int_{yy'} \Big[ \Gf_{k}^R(x,y) \,\partial_k \Rf_{k}^R(y,y') \, \Gf_{k}^K(y',x') \\ \nonumber
    &\hspace{0.8cm} +\;\Gf_{k}^K(x,y) \,\partial_k \Rf_{k}^A(y,y') \, \Gf_{k}^A(y',x') \\ \nonumber
    &\hspace{0.8cm} +\;\Gf_{k}^R(x,y) \,\partial_k \Rf_{k}^K(y,y') \, \Gf_{k}^A(y',x') \Big] \,.
\end{align}
\end{subequations}
In thermal equilibrium, the respective Keldysh components are also fixed by the FDR,
\begin{align}
    \Bb_{ab,k}^K(\omega, \vec{p})=\coth\left(\frac{\omega}{2T}\right)\left(\Bb_{ab,k}^R(\omega, \vec{p})-\Bb_{ab,k}^A(\omega, \vec{p})\right) \label{eq:BbFDR}
\end{align}
and
\begin{align}
    \Bf_k^K(\omega, \vec{p})=\tanh\left(\frac{\omega-\mu}{2T}\right)\left(\Bf_k^R(\omega, \vec{p})-\Bf_k^A(\omega, \vec{p})\right) \,. \label{eq:BfFDR}
\end{align}

With the flow equation \eqref{eq:floweq}
one can proceed to study the flow of the scale dependent effective potential which will allow one to compute the phase diagram.

\subsection{Flow equation of the effective potential}\label{eq:flowEqOfEffPot}
One can generally extract the effective potential $V_k(\phi)$ (with $\phi \equiv \phi^c/\sqrt{2}$) from the effective average Keldysh action $\Gamma_k$ via\footnote{To obtain the effective potential $V(\phi)$ from the effective (Keldysh) action $\Gamma$ one has to perform at least one functional derivative with respect to the quantum field $\phi^q$, since otherwise the partition function would be $Z = 1$ and the effective action hence identically equal to zero, $\Gamma = 0$.}
\begin{equation}
 -\frac{\partial V_k}{\partial \phi_a} \equiv \frac{1}{\sqrt{2}}\frac{\delta \Gamma_k}{\delta \phi^q_a(x)} \bigg\rvert_{\substack{\phi^c_a(x) \equiv \sqrt{2}\,\phi_a \,,\; \phi^q_a(x) \equiv 0\,,\\ \bar{\Psi}(x) \equiv 0  \,,\; \Psi(x)\equiv 0}} \,. \label{effPotDef}
\end{equation}
One can then use the flow equation Eq.~\eqref{eq:floweq} to compute the flow of the effective potential. Using \eqref{effPotDef}, upon Fourier transformation, the flow equation of $\partial V_k/\partial \phi_a$ can be written as
\begin{align}
    \label{dGamdPhiQFourierTransform}
    &-\sqrt{2}\, \partial_k\left( \frac{\partial V_k}{\partial \phi_a} \right) = \\ \nonumber
    &\hspace{1.0cm} i\int_{pq} \Big[
    \Gamma_k^{\bar{\Psi}_2\phi_a^q\Psi_1}(p,q,-p) \, \Bf_{k}^K(p) \\ \nonumber
    &+\Gamma_k^{\bar{\Psi}_1\phi_a^q\Psi_1}(p,q,-p) \, \Bf_{k}^R(p) +  \Gamma_k^{\bar{\Psi}_2\phi_a^q\Psi_2}(p,q,-p) \, \Bf_{k}^A(p) \Big] \\ \nonumber
    &\hspace{0.5cm} -\frac{i}{2} \int_{pq} \Big[ \Gamma_{k}^{\phi_b^c\phi_a^q \phi_c^c}(p,q,-p) \, \Bb_{cb,k}^F(p) \; + \\ 
    &\nonumber \Gamma_{k}^{\phi_b^q \phi_a^q \phi_c^c}(p,q,-p) \, \Bb_{cb,k}^R(p) +
    \Gamma_{k}^{\phi_b^c \phi_a^q \phi_c^q}(p,q,-p) \, \Bb_{cb,k}^A(p) \Big] \,,
\end{align}
where all quantities are understood in front of a constant and  purely classical bosonic field expectation value $\phi^c_a(x) \equiv \sqrt{2}\,\phi_a$.
Let us first look at the bosonic contribution in \eqref{dGamdPhiQFourierTransform}. Along the lines of Ref.~\cite{Roth:2024rbi} it can be shown that the bosonic contribution in \eqref{dGamdPhiQFourierTransform} can be formally integrated with respect to $\phi$ and that the result can be written as
\begin{align}
    &\partial_k V_k(\phi) = \frac{i}{4} \int_{\omega\vec{p}} \coth\left(\frac{\omega}{2T}\right)\times \\ \nonumber 
    &\left( \Gb_{ab,k}^R(\omega,\vec{p})  \,\partial_k \Rb_{ba,k}^R(\omega,\vec{p})  -\Gb_{ab,k}^A(\omega,\vec{p})  \,\partial_k \Rb_{ba,k}^A(\omega,\vec{p}) \right)  \,.
\end{align}
One can evaluate the frequency integral via the residue theorem. There one effectively picks up the poles corresponding to the bosonic Matsubara frequencies: For the part with the retarded propagator, one closes the contour in the upper half-plane so as to pick up the positive Matsubara modes. For the advanced propagator, one closes the contour in the lower half-plane to pick up the negative Matsubara frequencies. When the contour cross the pole at $\omega=0$, the residue contributes half of its value to the integral (this happens twice, however, for the retarded and for the advanced propagator). Then one can evaluate the frequency integral to obtain 
\begin{align}
\label{boson_contribution0}
    \partial_k V_k(\phi) &= \\ \nonumber
    &-\frac{T}{2} \int_{\vec{p}}\bigg[ \frac{1}{2}\big( \Gb^R_{ab,k}(0,\vec{p}) \, \partial_k \Rb_{ba,k}^R(0,\vec{p}) \;+ \\ \nonumber
    &\hskip 1.8cm \Gb^A_{ab,k}(0,\vec{p})  \,\partial_k \Rb_{ba,k}^A(0,\vec{p}) \big) \;+ \\ \nonumber
    &\sum_{n=1}^{\infty} \Gb_{ab,k}^R(i\omega_n, \vec{p}) \,\partial_k \Rb_{ba,k}^R(i\omega_n, \vec{p}) \;+ \\ \nonumber
    &\hskip .4cm \sum_{n=-\infty}^{-1}\! \Gb_{ab,k}^A(i\omega_n, \vec{p}) \,\partial_k \Rb_{ba,k}^A(i\omega_n, \vec{p}) \bigg]\,, 
\end{align}
with the bosonic Matsubara frequencies $\omega_n = 2\pi n T$. We now connect Eq.~\eqref{boson_contribution0} to its Euclidean counterpart.
The Euclidean propagator $\Gb_k^E$ is connected to the retarded/advanced propagators $\Gb_k^{R/A}$ from the Keldysh formalism through analytic continuation. For the retarded propagator, one has (where we have suppressed the $O(4)$ indices $a$ and $b$) 
\begin{subequations}
\begin{align}
    \Gb_{k}^{R}(\omega,\vec{p}) = \lim_{\epsilon\to 0} \Gb_{k}^E(\omega_E = -i(\omega + i\epsilon),\vec{p}) \,. \label{eq:ACFromGE2GR}
\end{align}
which is valid for all complex $\omega$ from the upper half-plane, i.e.~for any $\omega$ with $\Im \omega \geq 0$.
Similarly, for the advanced propagator one has
\begin{align}
    \Gb_{k}^{A}(\omega,\vec{p}) = \lim_{\epsilon\to 0} \Gb_{k}^E(\omega_E = -i(\omega - i\epsilon),\vec{p}) \,. \label{eq:ACFromGE2GA}
\end{align}\label{eq:ACFromGE2GRA}%
\end{subequations}
which is valid for all complex $\omega$ from the \emph{lower} half-plane, $\Im \omega \leq 0$.
Plugging the $n^{\text{th}}$ Matsubara frequency $\omega_E = \omega_n$ into \eqref{eq:ACFromGE2GRA}, we find
\begin{align}
    \Gb_k^E(\omega_n,\vec{p}) = \begin{cases}
        \Gb_k^R(i\omega_n,\vec{p}) &\text{for $n\geq 0$}\\
        \Gb_k^A(i\omega_n,\vec{p}) &\text{for $n\leq 0$}
    \end{cases} 
\end{align}
Assuming that the real-time regulator is chosen to comply with the causal structure of the Keldysh action (see Sec.~2.1 of Ref.~\cite{Roth:2023wbp}), the Euclidean regulator can be similarly related,
\begin{align}
    \Rb_k^E(\omega_n,\vec{p}) = -\begin{cases}
        \Rb_k^R(i\omega_n,\vec{p}) &\text{for $n\geq 0$}\\
        \Rb_k^A(i\omega_n,\vec{p}) &\text{for $n\leq 0$}
    \end{cases} 
\end{align}
With this, \eqref{boson_contribution0} becomes
\begin{align}
\label{boson_contribution}
    \partial_k V_k(\phi) &=
	\frac{T}{2} \sum_{n=-\infty}^{\infty} \int_{\vec{p}} \Gb_{ab,k}^E(\omega_n, \vec{p}) \,\partial_k \Rb_{ba,k}^E(\omega_n, \vec{p}) \,,
\end{align}
which concludes the bosonic contribution to the flow of the effective potential.

At high temperatures $T \gg \omega$ (or small frequencies) one can expand $\coth(\omega/2T)$ in a Taylor series and take the leading order, such that $\coth(\omega/2T) \approx 2T/\omega$ in the Rayleigh-Jeans limit. Note that in this case only the zeroth Matsubara mode contributes to the frequency integral, such that one is left with 
\begin{align}
\label{boson_contribution_cl}
    \partial_k V_k(\phi) &= 
	\frac{T}{2}\int_{\vec{p}} \Gb^E_{ab,k}(0,\vec{p})\partial_k \Rb_{ba,k}^E(0,\vec{p}) \,.
\end{align}
In this limit one can observe two things: First, at high temperatures $T$ all quantum fluctuations are effectively suppressed, such that the flow of the effective potential is purely generated by thermal fluctuations. Second, one can see that in this limit only the propagator at $\omega=0$ enters. Hence, any information about the time dependence of fluctuations drop out from the right-hand side of the flow equation. This is generally the case and simply means that dynamic properties of the system do not influence the static flow  in the classical limit (here explicitly verified for the flow of the effective potential). In short, the statics completely decouples from the dynamics at high temperatures $T$. This was already shown on more general grounds in Ref.~\cite{Roth:2024rbi} for the flow of the coarse-grained free energy in classical-statistical systems with reversible mode couplings. In the present work this means that at high temperatures the flow of the effective potential becomes independent of the damping coefficients in \eqref{eq:gamPiAndGamSigma}, since those only appear in the bosonic propagators for non-zero frequencies $\omega \neq 0$.

However, if we look at moderate to small temperatures, then all Matsubara modes contribute to the flow of the effective potential. In this case quantum fluctuations also contribute via the non-zero Matsubara frequencies. Although this may sound trivial, it can have important consequences: 
In the present work, for example, these contributions imply that the damping/kinetic coefficients \eqref{eq:gamPiAndGamSigma} in the advanced/retarded propagators at non-vanishing frequency generally also enter to the flow of the effective potential. Because of this, the dynamic properties of the system can be expected to have an effect on the thermodynamic grand potential as well. Then one can naturally ask the questions: will different dynamic properties (here for example the presence or absence of the bosonic damping in \eqref{eq:gamPiAndGamSigma}, or more generally dissipative versus diffusive dynamics with potential conservation laws and reversible mode couplings according to the different dynamic models) have an influence on the phase diagram of the theory, and if so, how large is this effect? For the quark-meson model with or without dissipation by meson damping, these questions will be addressed in Sec.~\ref{sct:results}. 

Now we turn to the fermionic contribution in the flow \eqref{dGamdPhiQFourierTransform} of the effective potential. We consider a truncation in which the vertices $\Gamma_k^{\bar{\Psi}_1\phi_a^q\Psi_1}=\Gamma_k^{\bar{\Psi}_2\phi_a^q\Psi_2} = g_k$ are given by an FRG scale $ k$ dependent but field and momentum independent Yukawa coupling $g_k$. In this case, the anomalous vertices $\Gamma_k^{\bar{\Psi}_1\phi_a^q\Psi_2}=\Gamma_k^{\bar{\Psi}_2\phi_a^q\Psi_1}=0$ vanish, and the only fermionic contribution to the flow of the effective potential is given by 
\begin{align}
	\label{dGamdPhiQFourierTransform2}
	-\sqrt{2}\,\partial_k \left(\frac{\partial V_k}{\partial \phi_a}\right)  &=
	i \,g_k \Tr\Big[ \int_{p} 
	 \Bf_{k}^K(p) \Big] \, ,
\end{align}
where $\Tr$ here stands for the trace over Dirac and color indices. One can use the fermionic FDR \eqref{eq:BfFDR} to write Eq.~\eqref{dGamdPhiQFourierTransform2} as
\begin{align}
	\label{dGamdPhiQFourierTransform3}
	-\sqrt{2}\,\partial_k \left(\frac{\partial V_k}{\partial \phi_a}\right)  &= \\ \nonumber
 &\hspace{-1.0cm} i\, g_k \Tr\Big[ \int_{p} 
	\tanh\left(\frac{\omega-\mu}{2T}\right) \left( \Bf_{k}^R(p)-\Bf_k^A(p) \right) \Big] \, .
\end{align}
The integral over frequency can be evaluated analytically using the residue theorem. To evaluate the integral over $p^0$ for the $\Bf^R_k(p)$ term, we close the contour in the upper half-plane. To evaluate the $p^0$ integral for the $\Bf^A_k(p)$ term, we close the integration contour in the lower half-plane. The result is
\begin{align}
	\label{dGamdPhiQFourierTransform4}
	&-\sqrt{2}\,\partial_k \left(\frac{\partial V_k}{\partial \phi_a}\right)  = \\ \nonumber
	&-i g_k \! \Tr\Big[ \int_{\vec{p}} \!\! 
	\big( 2T\sum_{n=0}^\infty \Bf_{k}^R(i\tilde{\omega}_n,\vec{p}) \! +\! 2T \! \! \sum_{n=-\infty}^{-1} \Bf_k^A(i\tilde{\omega}_n,\vec{p}) \big) \Big] \, ,
\end{align}
in which the frequencies in the retarded and advanced propagators are substituted by the fermionic Matsubara frequencies $i\tilde{\omega}_n = (2n+1)\pi Ti+\mu$ with the chemical potential $\mu$. Using the relations
\begin{subequations}
\begin{align}
    \frac{\delta \Gf_k^R}{\delta \phi_a} = \sqrt{2}\,\Gf_k^R g_k \Gf_k^R \,, \hspace{0.5cm} 
    \frac{\delta \Gf_k^A}{\delta \phi_a} =\sqrt{2}\,\Gf_k^A g_k \Gf_k^A \,,
\end{align}
\end{subequations}
Eq.~\eqref{dGamdPhiQFourierTransform4} can be integrated with respect to $\phi$,
\begin{align}
    \partial_k V_k(\phi) &=
	T\int_{\vec{p}} 
	\Tr \bigg[ \sum_{n=0}^{\infty} \Gf_{k}^R(i\tilde{\omega}_n, \vec{p})\partial_k \Rf_k^R(i\tilde{\omega}_n, \vec{p})\;+  \nonumber \\
    &\sum_{n=-\infty}^{-1} \Gf_k^A(i\tilde{\omega}_n, \vec{p})\partial_k \Rf_k^A(i\tilde{\omega}_n, \vec{p}) \bigg] \, .
\end{align}
Using the (fermionic) analytic continuation of the propagator,
\begin{align}
    \Gf_k^E(\tilde{\omega}_n,\vec{p}) = \begin{cases}
        \Gf_k^R(i\tilde{\omega}_n,\vec{p}) &\text{for $n\geq 0$}\\
        \Gf_k^A(i\tilde{\omega}_n,\vec{p}) &\text{for $n< 0$}
    \end{cases} 
\end{align}
and of the regulator (again assuming that it complies with the causal structure of the Keldysh action~\cite{kamenev_2011}),
\begin{align}
    \Rf_k^E(\tilde{\omega}_n,\vec{p}) = -\begin{cases}
        \Rf_k^R(i\tilde{\omega}_n,\vec{p}) &\text{for $n\geq 0$}\\
        \Rf_k^A(i\tilde{\omega}_n,\vec{p}) &\text{for $n< 0$}
    \end{cases} \,, 
\end{align}
we obtain
\begin{align}
    \partial_k V_k(\phi) &=
	-T \!\!
	\sum_{n=-\infty}^{\infty}  \int_{\vec{p}} \Tr\Big[ \Gf_{k}^E(\tilde{\omega}_n, \vec{p})\partial_k \Rf_k^E(\tilde{\omega}_n, \vec{p}) \Big] 
\end{align}
for the fermionic contribution to the flow.

For fermions, there is no zeroth Matsubara mode and hence all fermionic fluctuations are quantum. 
The leading fermionic contribution to the flow of the effective potential at asymptotically  high temperatures $T$ vanishes.

Combining fermionic and bosonic contribution, the flow of the effective potential is given by
\begin{align}
\label{boson_and_fermion_contribution}
    \partial_k V_k(\phi) =
	&\frac{T}{2} \sum_{n=-\infty}^{\infty} \int_{\vec{p}} \Gb_{ab,k}^E(\omega_n, \vec{p}) \,\partial_k \Rb_{ba,k}^E(\omega_n, \vec{p}) \\ \nonumber -&T
	\sum_{n=-\infty}^{\infty}  \int_{\vec{p}} \Tr\Big[ \Gf_{k}^E(\tilde{\omega}_n, \vec{p})\partial_k \Rf_k^E(\tilde{\omega}_n, \vec{p}) \Big] \,,
\end{align}
which precisely agrees with the standard Euclidean flow equation for the effective potential.

\subsection{Truncation}
\label{Truncation}

In order to carry out the remaining momentum integration and Matsubara sums, we need to specify our truncation and the regulators. In local potential approximation (LPA) only the  mesonic effective potential $U_k(\rho)$ is assumed to depend on the FRG scale~$k$, such that the truncated effective average action is given by 
\begin{widetext}
\begingroup 
\setlength\arraycolsep{3pt} 
\begin{align}
    \Gamma_k \!=\!  \int_x &\bigg[ (\bar{\Psi}_1, \bar{\Psi}_2)
    \!\begin{pmatrix}
        i\gamma^\mu\partial_\mu \!+\! i\epsilon u_\mu \gamma^\mu&  2i\epsilon u_\mu \gamma^\mu\tanh\frac{iu^\mu\partial_\mu-\mu}{2T}\\
     0 & i\gamma^\mu\partial_\mu \!-\! i\epsilon u_\mu \gamma^\mu
    \end{pmatrix}
    \!\begin{pmatrix}
        \Psi_1 \\ \Psi_2
    \end{pmatrix} 
     -\frac{g}{\sqrt{2}}
    (\bar{\Psi}_1,\bar{\Psi}_2)
    \!\begin{pmatrix}
        \sigma^c \!+\! i\gamma_5 \vec{\tau}\cdot\vec{\pi}^c & \sigma^q \!+\! i\gamma_5 \vec{\tau}\cdot\vec{\pi}^q \\
        \sigma^q \!+\! i\gamma_5 \vec{\tau}\cdot\vec{\pi}^q & \sigma^c \!+\! i\gamma_5 \vec{\tau}\cdot\vec{\pi}^c
    \end{pmatrix}
    \!\begin{pmatrix}
        \Psi_1 \\ \Psi_2
    \end{pmatrix} \nonumber \\
    & + \frac{1}{2}(\phi_a^c, \phi_a^q) \!
    \begin{pmatrix}
        0 & -\partial^2 \!+\! \gamma u^\mu\partial_\mu\\
    -\partial^2 \!-\! \gamma u^\mu\partial_\mu& -2\gamma u^\mu\partial_\mu \coth\frac{iu^\mu\partial_\mu}{2T}
    \end{pmatrix} \!
    \begin{pmatrix}
        \phi_a^c\\ \phi_a^q
    \end{pmatrix} 
    -\frac{1}{2}  \rho^q(x) \, 2Yu^{\mu}\partial_{\mu} \coth\left(\frac{iu^{\mu}\partial_{\mu}}{2T}\right) \, \rho^q(x) \nonumber \\
    &-\frac{1}{2} \rho^q(x) \, Y u^{\mu}\partial_{\mu} \, \big(\rho(x)+\rho^{qq}(x)\big) 
    -U_{k}(\rho+\rho^q+\rho^{qq}) + U_{k}(\rho-\rho^q+\rho^{qq}) + \sqrt{2}\,c\sigma^c\bigg] \,. \label{eq:effAvgAction}
\end{align}
\endgroup
\end{widetext}
In this truncation, the fermionic sector is kept to be the same as the bare action, but we couple the bosonic part to an external heat bath with (FRG-scale-independent) damping coefficients $\gamma$ and $Y$ as described in Sec.~\ref{sct:bosonicHeatBath}. The coefficient $\epsilon$ in the fermionic Green function is infinitesimal, $\epsilon \to 0^+$.
A non-vanishing chemical potential for fermions is included. Such a chemical potential enters through the Keldysh propagator as $\tanh((\omega-\mu)/2T)$, which ensures that the symmetry of thermal equilibrium is maintained at finite temperature and chemical potential. 

For the present work, we choose the optimized regulator \cite{Litim:2001up}
\begin{subequations}
\begin{align}
    \Rb_{ab,k}^R(\omega,\vec{p}) &= -(k^2-\vec{p}^2)\,\theta(k^2-\vec{p}^2) \, \delta_{ab} \,, \\
    \Rb_{ab,k}^A(\omega,\vec{p}) &= -(k^2-\vec{p}^2)\,\theta(k^2-\vec{p}^2) \, \delta_{ab} \,, \\
    \Rb_{ab,k}^K(\omega,\vec{p}) &= 0 \,.
\end{align}\label{eq:optRegB}%
\end{subequations}
for the bosons, and its fermionic extension
\begin{subequations}
\begin{align}
    \Rf_{k}^R(\omega,\vec{p}) &= -\vec{\gamma}\cdot \vec{p}\, \left(\sqrt{\frac{k^2}{\vec{p}^2}} -1\right)\,\theta(k^2-\vec{p}^2) \,, \\
    \Rf_{k}^A(\omega,\vec{p}) &= -\vec{\gamma}\cdot \vec{p}\, \left(\sqrt{\frac{k^2}{\vec{p}^2}} -1\right)\,\theta(k^2-\vec{p}^2) \,, \\
    \Rf_{k}^K(\omega,\vec{p}) &= 0 \,.
\end{align}\label{eq:optRegF}%
\end{subequations}
for the quarks. As such, we restrict ourselves to a frequency-independent regulator here. If one needs to use a frequency-dependent regulator for some reason, one can follow the general construction scheme introduced in Refs.~\cite{Roth:2021nrd,Roth:2023wbp}, where one imagines the regulator to represent a coupling to an FRG-scale-dependent fictitious heat bath. We expect that a similar construction scheme is also possible in the fermionic case. In the spirit of Ref.~\cite{Otto:2022jzl}, it would be interesting to try different frequency-dependent regulators and assess the changes  this induces in phase diagram from the perspective of the real-time formalism.

From the effective average action \eqref{eq:effAvgAction} one can extract the longitudinal (sigma) and transverse (pion) propagators. At constant classical field expectation value $\phi^c \neq 0$ and $\phi^q=0$ (with $\rho=\frac{1}{2}\phi^c_a \phi^c_a$), one can decompose the bosonic propagator into longitudinal and transverse components (in field space) via 
\begin{align}
    G^{X}_{ab,k}(\omega,\vec{p}) &= \\ \nonumber
    &\hspace{-0.5cm} G^{X}_{\sigma,k}(\omega,\vec{p})\frac{\phi_a^c\phi_b^c}{2\rho}+G^{X}_{\pi,k}(\omega,\vec{p})\left(\delta_{ab}-\frac{\phi_a^c\phi_b^c}{2\rho}\right) \,,
\end{align}
with $X = R,A,K$.
From \eqref{eq:effAvgAction}, we then obtain
\begin{widetext}
\begin{subequations}
\begin{align}
    \Gb^R_{\sigma,k}(\omega,\vec{p})&=\frac{-1}{\omega^2+i(\gamma+Y\rho)\omega-\vec{p}^2-m_{\sigma}^2 + \Rb_{k}^R(\omega,\vec{p})} \,, \label{eq:GRSigma} \\
    \Gb^A_{\sigma,k}(\omega,\vec{p})&=\frac{-1}{\omega^2-i(\gamma+Y\rho)\omega-\vec{p}^2-m_{\sigma}^2 + \Rb_{k}^A(\omega,\vec{p})} \,, \label{eq:GASigma} \\
    \Gb^K_{\sigma,k}(\omega,\vec{p})&=\coth\left(\frac{\omega}{2T}\right) \left(\Gb^R_{\sigma,k}(\omega,\vec{p})-\Gb^A_{\sigma,k}(\omega,\vec{p})\right) \label{eq:GKSigma}
\end{align}
\end{subequations}
for the longitudinal (sigma) propagators
with (scale dependent) Euclidean mass parameter $m_{\sigma}^2 \equiv 2U_k'(\rho)+4\rho U''_k(\rho)$, and
\begin{subequations}
\begin{align}
    \Gb^R_{\pi,k}(\omega,\vec{p})&=\frac{-1}{\omega^2+i\gamma\omega-\vec{p}^2-m_{\pi}^2+\Rb_{k}^R(\omega,\vec{p})} \,, \label{eq:GRPi}\\
    \Gb^A_{\pi,k}(\omega,\vec{p})&=\frac{-1}{\omega^2-i\gamma\omega-\vec{p}^2-m_{\pi}^2+\Rb_{k}^A(\omega,\vec{p})} \,, \label{eq:GAPi}\\
    \Gb^K_{\pi,k}(\omega,\vec{p})&=\coth\left(\frac{\omega}{2T}\right) \left(\Gb^R_{\pi,k}(\omega,\vec{p})-\Gb^A_{\pi,k}(\omega,\vec{p})\right) \label{eq:GKPi}
\end{align}
\end{subequations}
for the transverse (pion) propagators with $m_{\pi}^2 \equiv 2U_k'(\rho)$.
The fermionic propagators are given by
\begin{subequations}
\begin{align}
    \Gf_{k}^R(\omega,\vec{p}) &= -\left((\omega+i\epsilon)\gamma^0 - \vec{\gamma}\cdot\vec{p} - \Mf + \Rf_{k}^R(\omega,\vec{p}) \right)^{-1} \,, \\
    \Gf_{k}^A(\omega,\vec{p}) &= -\left((\omega-i\epsilon)\gamma^0 - \vec{\gamma}\cdot\vec{p} - \Mf + \Rf_{k}^A(\omega,\vec{p}) \right)^{-1} \,, \\
    \Gf_{k}^K(\omega,\vec{p}) &= \tanh\left(\frac{\omega-\mu}{2T}\right)\left( \Gf_k^R(\omega,\vec{p})- \Gf_k^A(\omega,\vec{p})\right) \,,
\end{align}
\end{subequations}
\end{widetext}
with $\Mf = \frac{g}{\sqrt{2}}(\sigma^c+i\gamma_5 \vec{\tau}\cdot\vec{\pi}^c)$.

With a frequency independent regulator one can carry out the Matsubara sums in \eqref{boson_and_fermion_contribution} analytically and, with the convenient form of the optimized regulator \eqref{eq:optRegB} and its fermionic extension \eqref{eq:optRegF}, also analytically perform the momentum integration, to obtain explicitly the LPA flow of the effective potential in the form  
\begin{align}
    \partial_k U_k(\rho) = \frac{1}{2} I_{\sigma} + \frac{3}{2} I_{\pi} -2N_c I_{\psi} \,, \label{eq:dUdkDamping}
\end{align}
with
\begin{subequations}
\begin{align}
    I_{\sigma} &= \frac{k^4}{3\pi^2} 
\bigg\{-\frac{T}{k^2+m_{\sigma}^2}  +\frac{i}{2\pi E_{\sigma}} \times \label{eq:IsigmaDamping} \\ \nonumber
&\hspace{0.2cm} \bigg[ \psi\bigg( 
\frac{\tfrac{i}{2} (\gamma+Y\rho) + E_{\sigma} }{2\pi i T} \bigg) - \psi\bigg( 
\frac{\tfrac{i}{2} (\gamma+Y\rho) - E_{\sigma} }{2\pi i T} \bigg) \bigg]\bigg\} \,,   \\
    I_{\pi} &= \frac{k^4}{3\pi^2} \bigg\{ 
-\frac{T}{k^2+m_{\pi}^2} + \frac{i}{2\pi E_{\pi}} \times  \label{eq:IpiDamping} \\ \nonumber
&\hspace{0.2cm} \bigg[ \psi\bigg( 
\frac{\tfrac{i}{2} \gamma + E_{\pi} }{2\pi i T} \bigg) - \psi\bigg( 
\frac{\tfrac{i}{2} \gamma - E_{\pi} }{2\pi i T} \bigg) \bigg] \bigg\} \,, \\
    I_{\psi} &= \frac{k^4}{6\pi^2} \frac{\tanh\left(\frac{E_{\psi}-\mu}{2T}\right) + \tanh\left(\frac{E_{\psi}+\mu}{2T}\right)}{E_{\psi}} \,, \label{eq:IpsiDamping}
\end{align}
\end{subequations}
where $\psi(x)$ is the digamma function, 
and the scale-dependent single-particle energies are given by
\begin{subequations}
\begin{align}
    E_{\sigma} &= \sqrt{k^2+m_{\sigma}^2 - \tfrac{1}{4}  \left(\gamma+Y\rho \right)^2} \,, \\
    E_{\pi} &= \sqrt{k^2+m_{\pi}^2 - \tfrac{1}{4}\gamma^2} \,, \\
    E_{\psi} &= \sqrt{k^2+m_{\psi}^2} \,,
\end{align} \label{eq:singleParticleEnergies}%
\end{subequations}
with the (scale-dependent) quark mass $m_{\psi}^2 \equiv g^2 \rho$.
For sufficiently small values of the damping constants, i.e. for  $\gamma_\sigma= \gamma+Y\rho < 2m_{\sigma}$ and/or $\gamma_\pi=\gamma < 2m_{\pi}$, the sigma and/or pion fluctuations are represented by weakly-damped quasiparticle excitations. Otherwise,  the corresponding single-particle energies in Eqs.~\eqref{eq:singleParticleEnergies} become pure imaginary at $k^2= (\gamma+Y\rho)^2/4 - m_{\sigma}^2 $ and/or $k^2 = \gamma^2/4 - m_{\pi}^2 $, respectively, during the FRG flow, which corresponds to the sigma and/or pion excitations turning purely relaxational in the overdamped case.

In the limit $\gamma\rightarrow 0$ and $Y\rightarrow 0$ the flow equation \eqref{eq:dUdkDamping} reduces to the standard flow \cite{Schaefer:2004en} of the effective potential in the quark-meson model, which can be shown using the identity of the digamma function,
\begin{equation}
    \psi(-z) - \psi(z)  = 1/z+\pi \cot(\pi z) \,.
\end{equation}
In particular, with this, we obtain
\begin{subequations}
\begin{align}
    I_{\sigma} &= \frac{k^4}{6\pi^2} \frac{\coth\left(\frac{E_{\sigma}}{2T}\right)}{E_{\sigma}} \,, \label{eq:flowSigmaGam0} \\
    I_{\pi} &= \frac{k^4}{6\pi^2} \frac{\coth\left(\frac{E_{\pi}}{2T}\right)}{E_{\pi}} \,, \label{eq:flowPiGam0} 
\end{align}
\end{subequations}
and the single-particle energies in \eqref{eq:singleParticleEnergies} reduce to
\begin{align}
    E_{\sigma} = \sqrt{k^2+m_{\sigma}^2}\,, \;\;\;
    E_{\pi} = \sqrt{k^2+m_{\pi}^2} \,.
\end{align}

Another interesting limit is sending the pion and/or sigma damping to infinity. On the one hand, if we send only $Y \to \infty$ to infinity, the (infinite) overdamping only affects the sigma, and the corresponding loop function $I_{\sigma}$ becomes (see Appendix~\ref{sct:infDampCalc} for a derivation)
\begin{equation}
    I_{\sigma} \to \frac{k^4}{3\pi^2} \frac{T}{k^2+m_{\sigma}^2} \,.\label{eq:sigmaInfDamp}
\end{equation}
On the other hand, if we send $\gamma \to \infty$ to infinity, both sigma and pions will be (infinitely) overdamped, and in addition to \eqref{eq:sigmaInfDamp} we then also have
\begin{align}
    I_{\pi} \to \frac{k^4}{3\pi^2} \frac{T}{k^2+m_{\pi}^2} \,. \label{eq:piInfDamp}
\end{align}
Hence, the infinite-damping limit corresponds to taking into account contributions only from the zeroth Matsubara mode in \eqref{boson_contribution}. This is precisely the classical limit, where all quantum fluctuations are suppressed and only thermal fluctuations survive, the latter being determined by the Rayleigh-Jeans distribution. And just as in any classical-statistical system \cite{Roth:2024rbi}, the bosonic flow becomes independent of the dynamics, in this limit.

Nevertheless, in the general case of non-vanishing and finite bosonic damping coefficients, we find that even though damping/dissipation is a dynamic property of the system, the damping coefficients indeed potentially affect the flow equation of the effective potential through the non-zero Matsubara modes. The question now is how large the effect is, quantitatively, and we now turn to address this in the next section.

\section{Results for phase diagram with dissipation}\label{sct:results}
With the flow equation of the effective potential, one  computes the sigma condensate as the minimum of the effective potential at $k=0$, which is the order parameter of the system, for different temperatures and chemical potentials to obtain the phase diagram. In the numerical solution, we reformulate the flow equation \eqref{eq:dUdkDamping} as an advection-diffusion equation for the derivative $U_k'(\rho)$ \cite{Grossi:2019urj}, and discretize the resulting continuity equation on a grid in field space using the upwind scheme described in~\cite{Ihssen:2023qaq}.
\subsection{Finite bosonic damping}
\label{sct:phaseDiagramsFiniteDamping}
\begin{figure*}[t]
    \centering
    \begin{minipage}{0.32\textwidth}
        \centering
        {(a) $Y=0$, $\gamma=0$}\par\medskip
        \includegraphics[width=\linewidth]{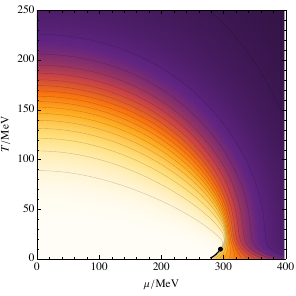}
        \end{minipage}
    \begin{minipage}{0.32\textwidth}
        \centering
        {(b) $Y \neq 0$, $\gamma=\gamma_1(T)$}\par\medskip
        \includegraphics[width=\linewidth]{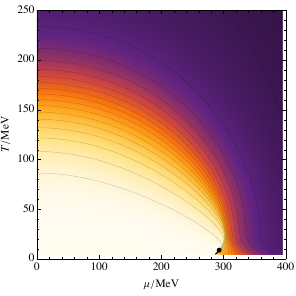}
        \end{minipage}
    \begin{minipage}{0.32\textwidth}
        \centering
        {(c) $Y\neq 0$, $\gamma=\gamma_2(T)$}\par\medskip
        \includegraphics[width=\linewidth]{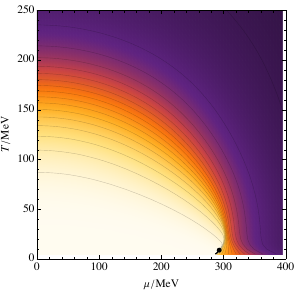}
    \end{minipage}
    \caption{Quark-meson model phase diagram  (a) for zero damping, (b) for constant $Y$ adjusted so that the sigma pole reproduces that of the broad two-pion resonance in the vacuum combined with the $O(4)$-symmetric damping $\gamma=\gamma_1(T)$ from Eq.~\eqref{eq:gamT5}, and (c) for $Y$ chosen as in (b) but with $\gamma=\gamma_2(T)$ from Eq.~\eqref{eq:gamT2}.}
    \label{fig:pdPhysDamp}
\end{figure*}

In LPA only the mesonic effective potential $U_k(\rho)$ is scale dependent, whereas especially the coefficients $\gamma$ and $Y$ that characterize the bosonic damping are fixed. One may also employ temperature and chemical-potential dependent values here, $\gamma=\gamma(\mu,T)$, $Y=Y(\mu,T)$. At the UV scale $k=\Lambda$, we start the flow with the parameterization 
\begin{equation}
U_\Lambda(\rho) = b_1 \rho+b_2 \rho^2 \label{eq:effPotUV}
\end{equation}
of the effective potential, with the values from Table~\ref{tab:UVInitCondsPhys}. We stop the flow in the infrared (IR) at a sufficiently small scale of $k_{\text{IR}}=40\,\mathrm{MeV}$, where we obtain the approximate values $\sigma_0 \approx 94$~MeV for the chiral order parameter (identified with the pion decay constant $f_{\pi}$ here), $m_{\psi} \approx 300$~MeV for the constituent quark mass, and $m_{\sigma} \approx 503$~MeV and $m_{\pi} \approx 136$~MeV for the Euclidean mass parameters of sigma meson and pions, respectively.
\begin{table}[t]
    \centering
    \begin{tabular}{cc|ccccc}
        $\gamma$ & $Y/\mathrm{MeV}^{-1}$ & $\Lambda/\mathrm{MeV}$ & $b_1/\mathrm{MeV}^2$ & $b_2$ & $c/\mathrm{MeV}^3$ & $g$ \\ \hline
        $0$ & $0$ & $1000$ & $3.15\cdot 10^5$ & $0.50$ & $1.75\cdot 10^6$ & $3.2$ \\
        Eq.~\eqref{eq:gamT5} & $0.064$ & $1000$ & $2.90\cdot 10^5$ & $0.86$ & $1.75\cdot 10^6$ & $3.2$ \\
        Eq.~\eqref{eq:gamT2} & $0.064$ & $1000$ & $2.90\cdot 10^5$ & $0.86$ & $1.75\cdot 10^6$ & $3.2$
    \end{tabular}
    \caption{UV initial conditions for vanishing (first row) and physically motivated  (second and third rows) pion and sigma damping constants, with $\gamma_{\pi}=\gamma$ and $\gamma_{\sigma}=\gamma + Y\rho$, cf.~Eq.~\eqref{eq:gamPiAndGamSigma}. The UV parameters are tuned such that the resulting vacuum observables are practically the same in the IR, for all cases.}
    \label{tab:UVInitCondsPhys}
\end{table}
We employ three different choices for the damping coefficients $\gamma$ and $Y$, and compare the resulting phase diagrams. The parameter set for the zero-damping case $\gamma=0$ and $Y=0$ corresponds to the one used in Ref.~\cite{Tripolt:2013jra}, and the resulting phase diagram can be seen in Fig.~\ref{fig:pdPhysDamp} (a). As a main feature, it has a first order phase transition at large chemical potentials and small temperatures, which terminates in a critical point at around $\mu_c \approx 293\,\mathrm{MeV}$ and $T_c \approx 10\,\mathrm{MeV}$.

To see what back-reaction a finite bosonic damping has on the phase diagram, we first turn on a non-zero value of $Y=0.064\,\mathrm{MeV}^{-1}$, which results in a sigma damping of $\gamma_{\sigma} = 556\,\mathrm{MeV}$ in the vacuum.\footnote{Note that at any fixed field expectation value $\rho$ the pair $(\gamma,Y)$ can be used interchangeably with $(\gamma_{\pi},\gamma_{\sigma})$, as the two pairs are related by $\gamma_{\pi}=\gamma$ and $\gamma_{\sigma}=\gamma+Y \rho$. If nothing else is specified, by $\gamma_{\sigma}$ we mean the sigma damping at the IR minimum $\rho = \sigma_0^2$.} 
With the width of the sigma given by $\gamma_\sigma/2$,   
this value yields the imaginary part of a sigma pole at approximately $(425 - i 278)\,\mathrm{MeV}$ which is  reasonably close to the corresponding physical two-pion resonance pole, 
e.g., see Ref.~\cite{Pelaez:2015qba}.

At finite temperature, the spectral function of the pion is known to be broadened by decay and capture processes from the medium \cite{Tripolt:2013jra}. To include these effects, we have to rely on external (phenomenological) input for the pion damping, since we do not solve for the damping self-consistently. At low temperatures, the behavior of the pion damping is known from chiral perturbation theory where, at leading order, one has $\gamma(T) = T^5/(12f_{\pi}^4)$ with $f_{\pi}=93$~MeV \cite{Goity:1989gs,Leutwyler:1990uq}.
To model the high-temperature behaviour of the pion damping, we follow two different approaches:
The results of Ref.~\cite{Lowdon:2022xcl} suggest a linear behaviour $\gamma(T) \sim T$ at high temperatures. Using the arguably simplest functional form that interpolates between the two behaviors at some matching temperature $T_{*}$, we parametrize
\begin{equation}
    \gamma_1(T) = \frac{T^5}{12f_{\pi}^4} \cdot \frac{1}{1+(T/T_{*})^4}  \label{eq:gamT5}
\end{equation}
for the pion damping as a function of temperature.
For the matching point $T_{*}$ we choose the pseudocritical temperature $T_{*} = T_{pc} \approx 175$~MeV. 
This constitutes our first model $\gamma_1(T)$ for the pion damping. 
(For the scope of this work, we neglect any dependence on chemical potential.) The proportionality constant for the linear behaviour $\gamma(T) \sim T$ at high $T \gg T_{*}$ is fixed to $T_{*}^4/(12f_{\pi}^4) \approx 1.04$ and yields, e.g.,~at $T = 220$~MeV to a pion damping of $\gamma \approx 164$~MeV.
Because the data of Ref.~\cite{Lowdon:2022xcl} suggests that the pion damping should be much larger at this temperature already, however, with a value of the  order $\sim 636$~MeV, we also  employ a second model for the pion damping, which reads
\begin{align}
    \gamma_2(T) = \frac{T^5}{12f_{\pi}^4} \big(1-F(T/T_{**})\big) + \alpha T \, F(T/T_{**}) \, , \label{eq:gamT2}
\end{align}
where  $F(x)$ represents  an interpolating function with $F(x) \to 1$ for $x \gg 1$, and $F(x) \to 0$  for $x \to 0$ faster than $x^4$ (so that the chiral perturbation theory result is not changed at low temperatures). $T_{**}$ again denotes a matching point between the two asymptotic behaviors, and $\alpha$ the proportionality constant at high temperatures. We use the arguably simplest interpolation function $F(x)=x^5/(1+x^5)$ that is consistent with these requirements, and choose the remaining constants $\alpha$ and $T_{**}$ by roughly matching the model \eqref{eq:gamT2} to the results from \cite{Lowdon:2022xcl}, which leads to $\alpha \approx 5.5$ and $T_{**} \approx 200$~MeV. This in total then constitutes our second model $\gamma_2(T)$ explored for the pion damping here. For more details and some caveats in this matching procedure, see App.~\ref{sec:DampingModels}.

The phase diagrams for the two models $\gamma_1(T)$ and $\gamma_2(T)$ are shown in Fig.~\ref{fig:pdPhysDamp} (b) and (c). One can see that there are only very minor quantitative changes in the phase diagram, when compared to the zero-damping case in (a).
\begin{figure*}
    \centering
    \begin{minipage}{0.49\textwidth}
        \centering
        {(a) $\mu = 0$}\par\medskip
        \includegraphics[width=\linewidth]{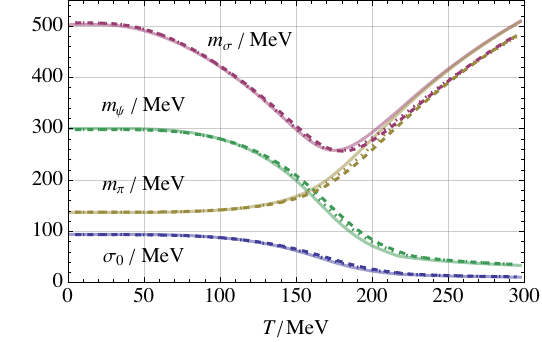}
    \end{minipage}
    \begin{minipage}{0.49\textwidth}
        \centering
        {(b) $T = 10\,\mathrm{MeV}$}\par\medskip
        \includegraphics[width=\linewidth]{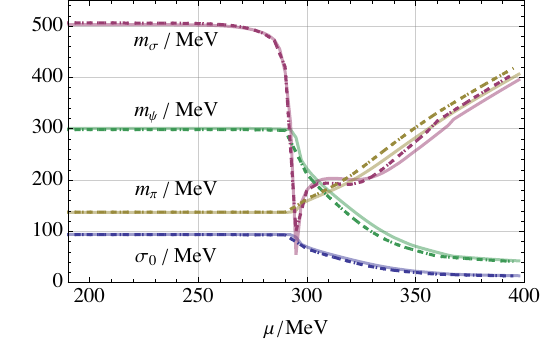}
    \end{minipage}
    \caption{Static IR observables that correspond to the three phase diagrams in Fig.~\ref{fig:pdPhysDamp}, for zero damping (solid lines), for the first model $\gamma=\gamma_1(T)$ defined in \eqref{eq:gamT5} (dotted), and for the second model $\gamma=\gamma_2(T)$ defined in \eqref{eq:gamT2} (dashed). As discussed in the main text, in both cases with $O(4)$-symmetric damping $\gamma \not= 0$ we use the same constant $Y$ in the sigma damping to reproduce the broad two-pion resonance pole in the vacuum.}
    \label{fig:plCompStaticObservablesPhysDamp}
\end{figure*}
In Fig.~\ref{fig:plCompStaticObservablesPhysDamp}, we plot the Euclidean (curvature) mass parameters in the IR, which are identical with the static screening masses in our present truncation, both as a function of temperature at $\mu=0$ and as a function of chemical potential at roughly the critical temperature, $T = 10\,\mathrm{MeV}$. The solid lines are the results for vanishing damping $\gamma=0$ and $Y=0$, and coincide with Fig.~4 of Ref.~\cite{Tripolt:2013jra}. The dotted lines correspond to $Y\neq 0$ and $\gamma=\gamma_1(T)$ modeled according to Eq.~\eqref{eq:gamT5}. The dashed lines correspond to $Y\neq 0$ and $\gamma=\gamma_2(T)$ modeled according to Eq.~\eqref{eq:gamT2}. For $\mu=0$ in Fig.~\ref{fig:plCompStaticObservablesPhysDamp}~(a), one observes that the finite damping starts having  an influence on the static observables essentially only at temperatures $T \gtrsim 150\,\mathrm{MeV}$, whereas at lower temperatures, for $T \lesssim 150\,\mathrm{MeV}$ where the pion damping becomes small anyway, the presence of the finite sigma damping has hardly any effect either.

Qualitatively, the finite damping causes the static observables (e.g.~the condensate $\sigma_0$) to lag behind their corresponding zero-damping values when the temperature is increased. In particular, the non-zero damping shifts the pseudocritical temperature (here defined as the minimum of $m_{\sigma}$) to  slightly larger values. This concurs with physical intuition, because a non-zero damping implies that bosonic fluctuations decay into the heat bath and, consequently, one needs higher temperatures for the same amount of fluctuations as compared to the case without damping, where all fluctuations are stable. 

Similarly small effects on the static screening masses are observed at $T=10$~MeV in their chemical potential dependence in Fig.~\ref{fig:plCompStaticObservablesPhysDamp}~(b), where modifications due to damping, if any, become noticeable also essentially only in the chirally restored regime, here for $\mu\gtrsim \mu_c$.  
In summary, although there is indeed some dependence of the phase diagram on the dynamics, consistent with our discussion in Sec.~\ref{eq:flowEqOfEffPot}, the quantitative effects on static observables are all rather minor for these physically motivated  (realistic) values of the sigma and pion damping.

\begin{figure*}
    \centering
    \begin{minipage}{0.49\textwidth}
        \centering
        {(a) $Y \neq 0$, $\gamma=\gamma_1(T)$}\par\medskip
        \includegraphics[width=\linewidth]{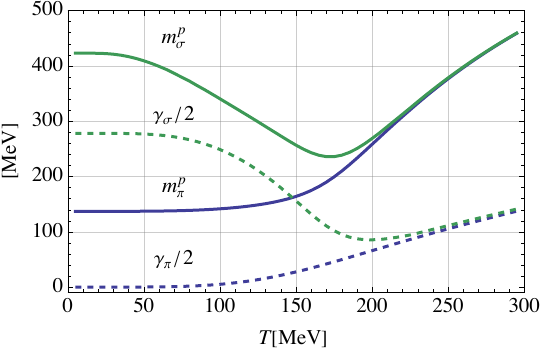}
    \end{minipage}
    \begin{minipage}{0.49\textwidth}
        \centering
        {(b) $Y \neq 0$, $\gamma=\gamma_2(T)$}\par\medskip
        \includegraphics[width=\linewidth]{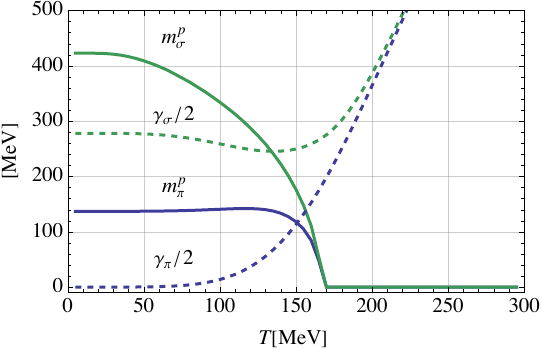}
    \end{minipage}
    \caption{Temperature dependence of real and imaginary parts of the sigma and pion poles at $\mu=0$, where (a) on the left and (b) on the right correspond to the two different models of pion damping $\gamma=\gamma_1(T)$ and $\gamma_2(T)$ as defined in Eqs.~\eqref{eq:gamT5} and \eqref{eq:gamT2}, respectively.}
    \label{fig:plPoleMasses}
\end{figure*}
In contrast, the effects of the two different models in  Eqs.~\eqref{eq:gamT5} and \eqref{eq:gamT2} for the (pseudo)scalar meson damping on dynamic quantities such as pole masses are much more pronounced. This can be seen in Fig.~\ref{fig:plPoleMasses}, where we plot both, the real and imaginary parts of the complex sigma and pion poles at  $\omega_{\sigma,\pi} = m_{\sigma,\pi}^p -i\gamma_{\sigma,\pi}/2$ (in the IR) as functions of temperature $T$ at $\mu=0$, for $Y \neq 0$, corresponding to a sigma damping that is always larger than that of the pion which vanishes in the vacuum. At finite temperature, the latter is given by the $O(4)$-symmetric damping $\gamma$, for which 
we compare the two physically motivated models: (a) with $\gamma=\gamma_1(T)$ according to Eq.~\eqref{eq:gamT5} in the left panel of Fig.~\ref{fig:plPoleMasses}, and (b) with $\gamma=\gamma_2(T)$ from  Eq.~\eqref{eq:gamT2} in the right panel of Fig.~\ref{fig:plPoleMasses}. With  $Y \neq 0$ in either case, we have $\gamma_{\sigma}-\gamma_{\pi} = Y\rho_{0}$, and the difference thus tends to zero as chiral symmetry gets restored, here at finite temperature (for $\mu=0$) in the crossover around the (static) pseudocritical temperature $T_{pc} \approx 175$~MeV.

Above this crossover, the finite but still relatively small $O(4)$-symmetric pion and sigma damping $\gamma$ from Eq.~\eqref{eq:gamT5}, relative to the real parts    of their pole masses in Fig.~\ref{fig:plPoleMasses}~(a), merely leads to a small correction of the latter as compared to the static high-temperature screening masses of pion and sigma in Fig.~\ref{fig:plCompStaticObservablesPhysDamp}~(a). 

This picture changes considerably, however, if a high-temperature pion damping is chosen of the order of magnitude as in the recent spectral reconstructions from lattice QCD data \cite{Lowdon:2022xcl}.
This is demonstrated in Fig.~\ref{fig:plPoleMasses}~(b) where we use our second model from Eq.~\eqref{eq:gamT2} for the temperature dependence of the pion damping in the chirally restored phase. The relatively large damping at high temperatures in this case causes the sigma and pion poles to become pure imaginary at temperatures above $T \gtrsim 170\,\mathrm{MeV}$. This implies that the sigma and pion excitations are no-longer propagating, but become purely relaxational  at these high temperatures and dissolve on time scales of the order of $1/\gamma_{\sigma,\pi}$ into the heat bath. It is interesting to note that the temperature of $T \approx 170$~MeV where this happens roughly matches the (static) pseudo-critical temperature $T_{pc} \approx 175$~MeV. In this spirit, our second model \eqref{eq:gamT2} can be seen to describe  a `deconfined' phase in which pions (and the sigma meson) no-longer exist as propagating quasiparticle excitations in the thermal medium. Nevertheless, even with a such qualitatively very different dynamic behavior as in the cases (a) and (b), especially at high temperatures, the effects on phase diagram and static observables such as the screening masses remain relatively minor, cf.~Figs.~\ref{fig:pdPhysDamp} and \ref{fig:plCompStaticObservablesPhysDamp}. This is an example of the difficulty  distinguishing different dynamics, such as the qualitatively different behavior of the excitations in Figs.~\ref{fig:plPoleMasses}~(a) and (b), by just looking at static observables such as those in Fig.~\ref{fig:plCompStaticObservablesPhysDamp} alone.

\subsection{Infinite damping}
\label{sct:phaseDiagramsInfiniteDamping}
To study the maximum effect of dissipation on the phase diagram that the couplings to the bosonic heat baths can have, we now investigate two different extreme cases of strong bosonic damping. First, we consider a strong but
purely $O(4)$-invariant heat-bath coupling, corresponding to the $Y \to \infty$ limit at $\gamma=0$, which results in an infinite damping for the sigma, via $\gamma_{\sigma}\to\infty$, but vanishing damping for the pions, $\gamma_{\pi}=0$. In the second example, we consider only the $O(4)$-vector coupling to the heat bath, sending $\gamma \to \infty$ at $Y=0$, which then corresponds to an infinite damping for both, sigma and pions, in an $O(4)$-symmetric way,  with $\gamma_{\sigma}=\gamma_{\pi}\to\infty$. We compare both of these limits again to the zero-damping $\gamma_{\sigma}=\gamma_{\pi}=0$ phase diagram. In all cases we tune to the initial UV parameters such that we obtain the same vacuum observables in the IR as in Sec.~\ref{sct:phaseDiagramsFiniteDamping}. Note that in the case of infinite damping for both the sigma and the pions, we had to lower the UV cutoff $\Lambda$ from $\Lambda=1000$~MeV to $\Lambda=647.3$~MeV in order to be able to obtain the same value for the Euclidean sigma mass parameter $m_{\sigma} \approx 503$~MeV as in Sec.~\ref{sct:phaseDiagramsFiniteDamping}. For a most direct comparison with the other two cases here, we then generally use this lower UV cutoff for all results in this subsection. The corresponding values for the UV initial conditions are given in Table~\ref{tab:UVInitCondInfDamp}.
\begin{table}[t]
    \centering
    \begin{tabular}{cc|ccccc}
        $\gamma$ & $Y$ & $\Lambda/\mathrm{MeV}$ & $b_1/\mathrm{MeV}^2$ & $b_2$ & $c/\mathrm{MeV}^3$ & $g$  \\ \hline
        $0$ & $0$ & $647.3$ & $0.56\cdot 10^5$ & $2.2$ & $1.75\cdot 10^6$ & $3.2$ \\
        $0$ & $\infty$ & $647.3$ & $0.72\cdot 10^5$ & $1.77$ & $1.75\cdot 10^6$ & $3.2$ \\
        $\infty$ & $0$ & $647.3$ & $1.12\cdot 10^5$ & $0.5$ & $1.75\cdot 10^6$ & $3.2$
    \end{tabular}
    \caption{UV initial conditions for infinite damping. The parameters are tuned such that the resulting vacuum observables in the IR are the same in all cases. In the third case ($\gamma_{\sigma}=\gamma_{\pi}\to\infty$) we had to lower the UV cutoff from $\Lambda=1000\,\mathrm{MeV}$ to $\Lambda=647.3\,\mathrm{MeV}$ in order to reproduce the value for $m_{\sigma} \approx 503$~MeV of Sec.~\ref{sct:phaseDiagramsFiniteDamping} (with $\Lambda=1000\,\mathrm{MeV}$). For a better comparison (especially at temperatures of order $2\pi T \sim \Lambda$) we have adjusted the UV cutoff to this lower value also in the other two cases here.}
    \label{tab:UVInitCondInfDamp}
\end{table}

\begin{figure*}[t]
    \centering
    \begin{minipage}{0.32\textwidth}
        \centering
        {(a) $\gamma_{\sigma}=\gamma_{\pi}=0$}
        \includegraphics[width=\linewidth]{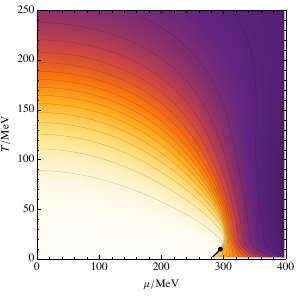}
        \end{minipage}
    \begin{minipage}{0.32\textwidth}
        \centering
        {(b) $\gamma_{\sigma}=\infty$, $\gamma_{\pi}=0$}
        \includegraphics[width=\linewidth]{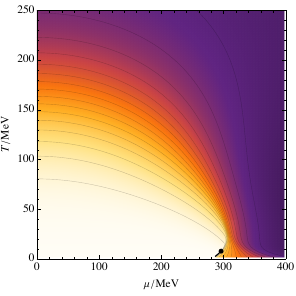}
    \end{minipage}
    \begin{minipage}{0.32\textwidth}
        \centering
        {(c) $\gamma_{\sigma}=\gamma_{\pi}=\infty$}
        \includegraphics[width=\linewidth]{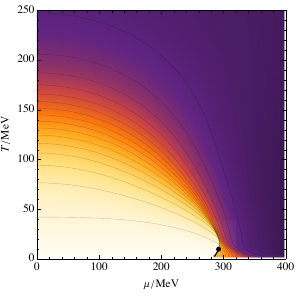}
    \end{minipage}
    \caption{Phase diagrams for (a) zero damping, (b) infinite damping for the sigma, but zero damping for the pion, and (c) infinite damping for both sigma and pions. Here the UV cutoff is reduced to $\Lambda = 647.3$~MeV, which quantitatively changes also the zero-damping phase diagram in (a) compared to the one in Fig.~\ref{fig:pdPhysDamp} (a) at high temperatures $T \sim \Lambda/(2\pi) \sim 103$~MeV.}
    \label{fig:pdsInfDamp}
\end{figure*}

\begin{figure*}
    \centering
    \begin{minipage}{0.49\textwidth}
        \centering
        {(a) $\mu = 0$}\par\medskip
        \includegraphics[width=\linewidth]{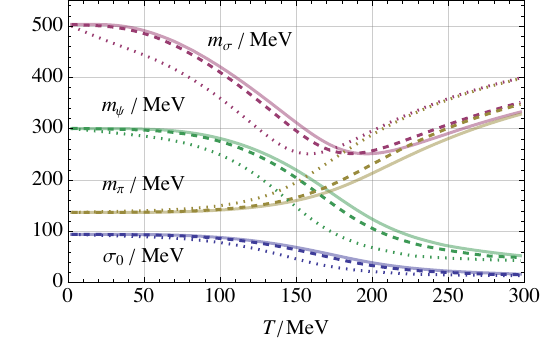}
    \end{minipage}
    \begin{minipage}{0.49\textwidth}
        \centering
        {(b) $T = 10\,\mathrm{MeV}$}\par\medskip
        \includegraphics[width=\linewidth]{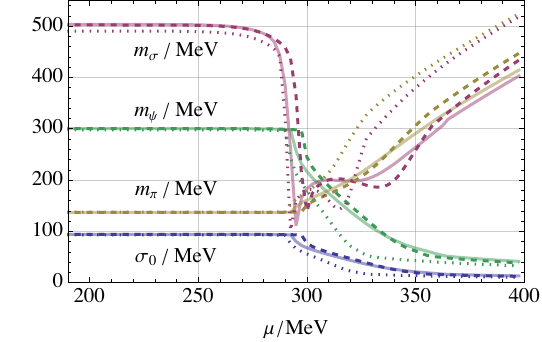}
    \end{minipage}
    \caption{Static IR observables for $\gamma_{\sigma}=\gamma_{\pi}=0$ (solid), $\gamma_{\sigma}=\infty$, $\gamma_{\pi} = 0$ (dashed) and $\gamma_{\sigma}=\gamma_{\pi}=\infty$ (dotted). Here the UV cutoff is $\Lambda = 647.3\,\mathrm{MeV}$.}
    \label{fig:plCompStaticObservablesInfDamp}
\end{figure*}

The resulting phase diagrams for the three cases are shown in Fig.~\ref{fig:pdsInfDamp}.
First, the zero-damping case, plotted in Fig.~\ref{fig:pdsInfDamp} (a), yields essentially the same phase diagram as in Fig.~\ref{fig:pdPhysDamp} (a), here with the smaller UV cutoff $\Lambda=647.3$~MeV, however. It is therefore expectable that some quantitative differences can occur for temperature of the order of  $T \sim \Lambda/(2\pi) \sim 103$~MeV. These quantitative differences are visible when comparing the solid lines from Fig.~\ref{fig:plCompStaticObservablesPhysDamp}~(a) and Fig.~\ref{fig:plCompStaticObservablesInfDamp} (a) for temperatures $T \gg 103$~MeV.
While not our main focus here,
such a residual dependence on the UV cutoff can in principle be addressed by requiring RG consistency of the resulting IR effective action \cite{Braun:2018svj}. 

Continuing with the second case in Fig.~\ref{fig:pdsInfDamp}~(b), where only the sigma damping is infinite, $\gamma_{\sigma}\to\infty$, but the pion damping vanishes, $\gamma_{\pi}=0$, we can conclude that the phase diagram remains qualitatively unchanged with some quantitative changes which are nevertheless comparatively rather small. Still qualitatively unchanged but with somewhat more pronounced quantitative changes the phase diagram for the third case, in which both damping constants are infinite, with $\gamma_{\sigma}=\gamma_{\pi}\to \infty$, is shown in Fig.~\ref{fig:pdsInfDamp}~(c).

In order to assess these modifications in some more detail, either from maximally strong overdamping of sigma alone or of pion and sigma excitations together, in Fig.~\ref{fig:plCompStaticObservablesInfDamp} we plot the same static IR observables as in Fig.~\ref{fig:plCompStaticObservablesPhysDamp}, again both, as a function of temperature at $\mu=0$, and as a function of chemical potential at $T=10$~MeV.  We see once more that even quantitatively the changes compared to the zero-damping case (solid lines) are still rather small in the strong sigma-damping limit (dashed) and somewhat more pronounced for strong $O(4)$-symmetric damping of sigma and pions (dotted).
One remarkable feature, which is also more prominent in the $\gamma_{\sigma}=\gamma_{\pi}\to\infty$ case, is the shift of the pseudocritical temperature to smaller values. Physically, this can be understood by recalling that in the infinite-damping limit, the bosonic contributions  to the flow equation in Eqs.~\eqref{eq:sigmaInfDamp} and  \eqref{eq:piInfDamp} become classical. Hence, the strength of thermal fluctuations of frequency $\omega$ is given by the Rayleigh-Jeans distribution $T/\omega$ instead of the Bose-Einstein distribution $1/(e^{\omega/T}-1)$. The former is always bigger than the latter, since quantum-mechanically, degrees of freedom can `freeze out' if their frequency is much higher than the temperature. Classically, however, the equipartition theorem gives all degrees of freedom an average kinetic energy of $\tfrac{3}{2}\,T$ (with $k_B=1$), independent of frequency. Hence, in the classical limit, it takes a smaller temperature for sufficiently strong  bosonic fluctuations to restore chiral symmetry, as compared to the quantum mechanical case. 

Note the non-monotonous behavior of the finite temperature transitions here: In the physically motivated damping models of the previous subsection the effect of dissipation of fluctuations dominated to delay chiral symmetry restoration, while in both strong-overdamping limits discussed here, in Fig.~\ref{fig:plCompStaticObservablesInfDamp}~(a), the enhanced fluctuations of the classical Rayleigh-Jeans limit apparently overcompensate this general trend of dissipation to cause the opposite overall effect.

While the chiral order parameter $\sigma_0$ melts more quickly with temperature in both overdamped cases, in line with the physical argument above, the behavior of its low-temperature dependence on the chemical potential in Fig.~\ref{fig:plCompStaticObservablesInfDamp}~(b) is more complicated.
While the critical chemical potential $\mu_c \sim 280$~MeV of the first-order transition for $T\to 0$ still roughly matches the zero-damping case in both strong-overdamping limits, the shape of the first-order line with increasing temperatures $0<T<T_c$  and  the location of the critical point at $T=T_c$ change.
For $\gamma_\sigma\to\infty$, $\gamma_\pi=0$, the temperature of $T=10$~MeV Fig.~\ref{fig:plCompStaticObservablesInfDamp}~(b) is just in the crossover region above $T_c$ already, but due to the backbending of the contour lines of Fig.~\ref{fig:pdsInfDamp}~(b) in this region the corresponding chiral transition (dashed lines) in  Fig.~\ref{fig:plCompStaticObservablesInfDamp}~(b) is shifted to larger chemical potentials. With the $O(4)$-symmetric overdamping of sigma and pions, $\gamma_{\sigma}=\gamma_{\pi} \to \infty$, on the other hand, this backbending in Fig.~\ref{fig:pdsInfDamp}~(c) is reduced, and the corresponding chiral transition (dotted lines)  in  Fig.~\ref{fig:plCompStaticObservablesInfDamp}~(b) occurs at a somewhat smaller chemical potential already. To a similar effect, at a fixed finite temperature $T> 0$, it takes smaller chemical potentials to reach the same amount of chiral-symmetry restoration, as compared to the zero damping, in this case as well. This is evident from Figs.~\ref{fig:pdsInfDamp}~(a) and (c), and can also be seen by comparing the solid ($\gamma_{\sigma}=\gamma_{\pi}=0$) and the dotted ($\gamma_{\sigma}=\gamma_{\pi} \to \infty$) lines for chemical potentials $\mu \gtrsim 300$~MeV in Fig.~\ref{fig:plCompStaticObservablesInfDamp}~(b).

\section{Conclusion and Outlook}
\label{sec:conclusion}

In this paper, we have studied the real-time dynamics of the quark-meson model.
In a first step, in Sec.~\ref{formalism} we have presented the general real-time formalism of the quark-meson model on the Schwinger-Keldysh closed-time path. In particular, in Sec.~\ref{sct:QMModelSKContour} we have constructed the Keldysh action for relativistic Dirac fermions by using the Larkin-Ovchinnikov convention~\cite{kamenev_2011} in the fermionic Keldysh rotation. Based on the discrete symmetry of thermal equilibrium previously known for bosons~\cite{Sieberer:2015hba} and for non-relativistic fermions~\cite{Altland:2020lbb}, we have formulated a symmetry of thermal equilibrium for relativistic Dirac fermions in Sec.~\ref{sct:symmThEq}. This symmetry expresses the invariance of the Keldysh action (and hence of correlation functions) under a combination of a Kubo-Martin-Schwinger (KMS) transformation with time reversal, and leads to fermionic fluctuation-dissipation relations (FDR) between real-time correlation functions.

With our real-time formulation of the quark-meson model at hand, we have then continued  in Sec.~\ref{Truncation} with incorporating bosonic dissipation (damping) by coupling the system to an external heat bath with Ohmic spectral distribution. To distinguish between pion and sigma damping in an $O(4)$ invariant way, we have introduced two independent sets of Gaussian degrees of freedom in the bath, where one set consists of an ensemble of $O(4)$ vectors $\chi_{s,a}$ and couples linearly via $\chi_{s,a}\phi_a$ to the system vector $\phi = (\sigma,\vec\pi) $ of sigma meson and pions, and the other one represents an ensemble of $O(4)$ scalars $\varphi_s$ which couple to the $O(4)$ field invariant $\rho = \phi_a \phi_a$ via $\varphi_s \rho$. Integrating out both of these oscillator sets and choosing Ohmic spectral distributions as in Eqs.~\eqref{eq:ohmicJgammaY} then gives rise to distinct pion and sigma damping terms. In particular, in this approach the difference between the two is related to the field expectation value, $\gamma_{\sigma}-\gamma_{\pi} \sim \rho$, and hence to the realization of chiral symmetry. 

In Sec.~\ref{sec:floweq} we have re-derived the fermionic part of the FRG flow equation within the real-time formalism. Since we start from the real-time theory, analytic continuation to the Euclidean domain is well defined and unique. We have shown explicitly, that via such an analytic continuation the standard Euclidean flow equation \cite{Schaefer:2004en} for the effective potential is recovered in the real-time approach. In particular, we have demonstrated that (unlike in classical systems), the flow equation of the effective potential can in principle depend on the \emph{dynamic} properties through the contributions from non-zero Matsubara modes. The latter generally contain indirect information on the time-dependence of the excitations in the system, and hence about the dynamics. This implies that, in principle, static properties (e.g.~the phase diagram and static screening masses) can depend on the dynamic properties. This is unlike classical systems, where only the zeroth Matsubara modes contribute to static observables, which can be used to prove that the static properties are independent of the dynamics \cite{Roth:2024rbi}. 

In Sec.~\ref{sct:results} we have considered two different QCD-inspired models for the temperature dependence of the pion and sigma damping, to exemplify this effect and to investigate its quantitative extent. Both models are based on the leading-order result from chiral perturbation theory \cite{Goity:1989gs,Leutwyler:1990uq} at low temperatures, with $\gamma_\pi \sim T^5 $,  which is combined by interpolation with linear temperature dependencies at asymptotically high temperatures, where  $\gamma_\sigma = \gamma_\pi \sim T $, of different strengths. With these physically motivated choices for the mesonic dampings, we have found the quantitative effects on the phase diagram to be rather minor. In contrast, however, our two distinct models have served to demonstrate how dynamic properties such as the pole masses can be drastically different for these choices at the same time. In particular, we have found that in our second model,  as motivated from reconstructions of pion spectral functions  at high temperatures from lattice-QCD data performed in Ref.~\cite{Lowdon:2022xcl}, the poles in the retarded sigma and pion propagators become pure imaginary describing purely relaxational meson excitations at temperatures roughly right above the pseudocritical temperature. 

To estimate the largest possible effect of bosonic damping we have also considered two limits of maximally strong overdamping,
where either only the (longitudinal) sigma damping or the $O(4)$-symmetric damping for sigma and pion are sent to infinity. Even in these extreme cases, the quantitative effects on the phase diagram, although more pronounced than for  finite damping, still remain comparatively small. Interestingly, however, in the case of maximally strong overdamping of sigma and pions, with $\gamma_\sigma=\gamma_\pi\to\infty$,  the bosonic flow  becomes purely classical, corresponding to the Rayleigh-Jeans limit of the Bose-Einstein distribution. In particular, this also implies that at strictly zero temperature all bosonic fluctuations (both thermal and quantum) are suppressed, such that 
the corresponding purely fermionic flow of a so-called extended mean-field approximation becomes exact in this limit. 

In the future, the present work can be extended in various directions. First, for a somewhat more fundamental description of the environment, one might promote the non-relativistic Ohmic dampings in Eqs.~\eqref{eq:ohmicJgammaY} by a relativistic formulation in terms of Gaussian ensembles of Klein-Gordon fields with spectral distributions in terms of the invariant $s=\omega^2-\vec{p}^2$.
In such a formulation, the distinction between space and time arises only from the thermal density matrix, i.e.~from the bosonic and fermionic distribution functions in the Keldysh component of the self energies, cf.~Eq.~\eqref{eq:keldyshActionWithEps}, whereas the retarded and advanced components of the self-energy would be fully Lorentz-invariant (at least at tree level).

Another direction one could follow would be to introduce dissipation also for the fermions, and to investigate the corresponding effects on the phase diagram. Here one has to be more careful, since in order for the chemical potential to remain well defined, the dissipation must not violate the conservation of the Noether current associated with the $U(1)_B$ symmetry. Because of this, dissipation in the fermionic sector should rather be described in terms of baryon-charge diffusion. 
This would require coupling conserved baryon-number current $j^\mu = \bar{\psi}\gamma^\mu \psi $ to an ensemble of equally conserved vector Hubbard fields. 
At finite density and quark masses, density fluctuations are well known to mix with those of the chiral condensate, e.g., see~\cite{Haensch:2023sig}. 
In presence of this mixing the critical mode becomes a linear combination of both \cite{Son:2004iv}, so that a proper inclusion of the diffusion of baryon charge would allow to incorporate at least Model-B dynamics in the quark-meson model.\footnote{Model~B is interesting because it can be seen as a precursor for the more intricate Model~H, which is widely assumed to be the dynamic universality class of the QCD critical point \cite{Son:2004iv}. Starting from Model~B, Model~H could be realized by a reversible coupling of the conserved order parameter to the diffusive shear modes of the energy-momentum tensor in a next step.}  One could then go beyond the universal properties of Model~B, and also estimate non-universal corrections (which are present at any finite `distance' to the critical point) within an effective description of the different realizations of chiral symmetry in dense QCD matter and the transitions between them, as a first step towards assessing to which extend critical dynamics can be observed near the QCD critical point.

\acknowledgments
We thank Peter Lowdon, Ugo Mire, Jan Pawlowski, Fabian Rennecke, and Jonas Wessely for insightful discussions. This work was supported by the Deutsche Forschungsgemeinschaft (DFG, German Research Foundation) through the CRC-TR 211 ‘Strong-interaction matter under extreme conditions’-project number 315477589 – TRR 211. JVR is supported by the Studienstiftung des deutschen Volkes. 

\appendix

\section{Effective Keldysh action}
\label{sct:EKA}

Introducing external sources for the bosonic and fermionic degrees of freedom in the partition function \eqref{eq:partFnc}, one obtains the generating functional for real-time correlation functions,
\begin{align}
    &Z[\eta_1,\bar{\eta}_1,\eta_2,\bar{\eta}_2,j^q,j^c] \equiv \label{eq:Z} \\ \nonumber
    &\hspace{0.2cm} \int \mathcal{D}\bar{\psi}_1 \mathcal{D}\psi_1 \mathcal{D}\bar{\psi}_2 \mathcal{D}\psi_2 \mathcal{D}\phi^c \mathcal{D}\phi^q \, \exp\bigg\{ iS \;+ \\ \nonumber
    &\hspace{0.5cm} i\int_x \left[ j_a^q \phi_a^c + j_a^c \phi_a^q + \bar{\eta}_1\psi_2 + \bar{\psi}_2\eta_1 + \bar{\eta}_2\psi_1 + \bar{\psi}_1\eta_2 \right] \bigg\} \,.
\end{align}
Its logarithm, the Schwinger functional $W$, is the generator of connected correlation functions,
\begin{equation}
    W[\eta_1,\bar{\eta}_1,\eta_2,\bar{\eta}_2,j^q,j^c] \equiv -i\log Z[\eta_1,\bar{\eta}_1,\eta_2,\bar{\eta}_2,j^q,j^c] \,. \label{eq:W}
\end{equation}
One-point functions (i.e.~field expectation values) are obtained from $W$ via functional derivatives
\begin{subequations}
\begin{align}
    \phi_a^c &= \frac{\delta W}{\delta j_a^q} \,, \\
    \phi_a^q &= \frac{\delta W}{\delta j_a^c} \,, \\
    \psi_1 &= \frac{\delta W}{\delta \bar{\eta}_2} \equiv \frac{\overrightarrow{\delta}}{\delta \bar{\eta}_2} W \,, \\
    \psi_2 &= \frac{\delta W}{\delta \bar{\eta}_1} \equiv \frac{\overrightarrow{\delta}}{\delta \bar{\eta}_1} W \,, \\
    \bar{\psi}_1 &= \frac{\delta W}{\delta \bar{\eta}_2} \equiv W \frac{\overleftarrow{\delta}}{\delta \eta_2} \,, \\
    \bar{\psi}_2 &= \frac{\delta W}{\delta \eta_1} \equiv W \frac{\overleftarrow{\delta}}{\delta \eta_1} \,,
\end{align} \label{eq:fieldExpValsFromWDerivs}%
\end{subequations}
where we use the convention that derivatives with respect to the bar fields $\bar{\psi}, \bar{\eta}, \ldots$ act from the left, and derivatives with respect to to $\psi,\eta,\ldots$ act from the right.

The effective action $\Gamma$, the generator of one-particle irreducible (1PI) vertex functions, is related to the Schwinger functional via a Legendre transform,
\begin{align}
    &\Gamma[\bar{\psi}_1,\psi_1,\bar{\psi}_2,\psi_2,\phi^c,\phi^q] \equiv W \;- \label{eq:Gam} \\ \nonumber
    &\hspace{1.0cm} \int_x \big[ j_a^q \phi_a^c + j_a^c \phi_a^q + \bar{\eta}_1\psi_2 + \bar{\psi}_2\eta_1 + \bar{\eta}_2\psi_1 + \bar{\psi}_1\eta_2 \big] \,,
\end{align}
where the sources on the right-hand side are implicitly given by inverting the relations in \eqref{eq:fieldExpValsFromWDerivs}.

\section{Infinite damping calculation}
\label{sct:infDampCalc}

Let $I$ be a bosonic contribution to the flow of the form
\begin{align}
    I &= \frac{k^4}{3\pi^2}\bigg\{ -\frac{T}{E^2+\gamma^2/4} + \frac{i}{2\pi E} \times \label{eq:genBosonicFlow} \\ \nonumber
    &\hspace{2.5cm} \bigg[ \psi\bigg( 
\frac{i \gamma/2 + E }{2\pi i T} \bigg) - \psi\bigg( 
\frac{i \gamma/2 - E }{2\pi i T} \bigg) \bigg] \bigg\} 
\end{align}
with $E = \sqrt{M^2 - \gamma^2/4}$ and $M^2>0$.
In the limit $\gamma\to \infty$, with $M^2$ fixed, the energy $E$ behaves as
\begin{equation}
    E = \frac{i\gamma}{2} - \frac{iM^2}{\gamma} +\cdots
\end{equation}
such that the arguments of the digamma functions in \eqref{eq:genBosonicFlow} become
\begin{align}
\frac{i \gamma/2 + E }{2\pi i T} &=  
\frac{i \gamma - iM^2/\gamma + \cdots }{2\pi i T}  = 
\frac{\gamma}{2\pi T} + \cdots \\
\frac{i \gamma/2 - E }{2\pi i T}&=  
\frac{iM^2/\gamma + \cdots }{2\pi i T}= 
\frac{M^2}{2\pi \gamma T} +\cdots
\end{align}
To evaluate these further, we use the asymptotic expansions
\begin{alignat}{2}
    \psi(z) &= \log z + \order{\frac{1}{z}} \hspace{0.5cm} && (z\to\infty) \\
    \psi(z) &= -\gamma_E - \frac{1}{z} + \order{z} \hspace{0.5cm} && (z\to 0)
\end{alignat}
where $\gamma_E$ is the Euler-Mascheroni constant. This yields
\begin{align}
    \psi\left( \frac{\gamma}{2\pi T}+\cdots \right) &= \log\left( \frac{\gamma}{2\pi T} \right) + \cdots \\
    \psi\left(  \frac{M^2}{2\pi \gamma T}+\cdots \right) &= -\gamma_E - \frac{2\pi \gamma T}{M^2} + \cdots
\end{align}
With this, \eqref{eq:genBosonicFlow} becomes
\begin{align}
    I &= \frac{k^4}{3\pi^2}\bigg\{ -\frac{T}{M^2} + \frac{i}{2\pi E} \times \label{eq:genBosonicFlowExp} \\ \nonumber
    &\hspace{2.5cm} \left[ \log\left( \frac{\gamma}{2\pi T} \right)  + \gamma_E + \frac{2\pi \gamma T}{M^2} + \cdots \right] \bigg\} 
\end{align}
where we also used that $E^2+\gamma^2/4 = M^2$ is independent of $\gamma$. Expanding
\begin{equation}
    \frac{i}{2\pi E} = \frac{1}{\pi \gamma} + \cdots
\end{equation}
further yields
\begin{align}
    I &= \frac{k^4}{3\pi^2}\bigg\{ -\frac{T}{M^2} + \frac{1}{\pi \gamma} \times  \label{eq:genBosonicFlowExp2} \\ \nonumber
    &\hspace{2.5cm} \left[ \log\left( \frac{\gamma}{2\pi T} \right)  + \gamma_E + \frac{2\pi \gamma T}{M^2} \right] + \cdots \bigg\}
\end{align}
We see that in the infinite-damping limit $\gamma\to \infty$ both the $(1/\gamma)\log(\gamma/2\pi T)$ and the $\gamma_E/\gamma$ terms vanish, such that only the one survives where the two $\gamma$'s in the numerator and the denominator cancel, which finally yields
\begin{equation}
    I \to \frac{k^4 T}{3\pi^2 M^2} \hspace{1.0cm}(\gamma \to \infty) \label{eq:genBosonicFlowExp3}
\end{equation}

\section{Temperature dependence of the pion damping}
\label{sec:DampingModels}

\begin{figure}[t]
    \centering
    \includegraphics[width=\linewidth]{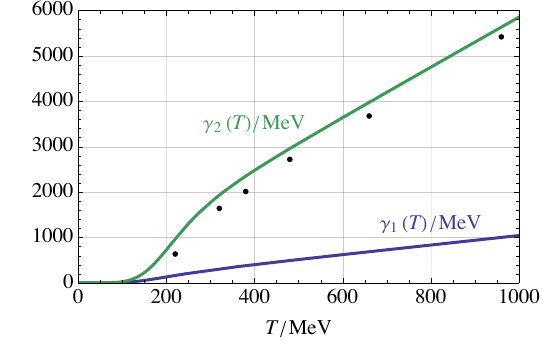}
    \caption{The two models $\gamma_1(T)$ (blue) and $\gamma_2(T)$ (green) for the temperature dependence of the pion damping from Eqs.~\eqref{eq:gamT5} and \eqref{eq:gamT2} plotted against the reconstructed damping-factor exponents of Ref.~\cite{Lowdon:2022xcl} (black). }
    \label{fig:DampingModels}
\end{figure}

The two models $\gamma_1(T)$ and $\gamma_2(T)$ for the temperature dependence of the pion damping defined in Eqs.~\eqref{eq:gamT5} and \eqref{eq:gamT2} are plotted in Fig.~\ref{fig:DampingModels} against the damping factor exponents of Ref.~\cite{Lowdon:2022xcl} that were obtained using spectral reconstruction of lattice QCD data. The trend of the black data points in Fig.~\ref{fig:DampingModels} suggests a linear dependence of the damping factor exponents at high temperatures, which is incorporated in both models $\gamma_1(T)$ and $\gamma_2(T)$. However, one can see in Fig.~\ref{fig:DampingModels} that the first model $\gamma_1(T)$ underestimates the size of the pion damping at higher temperatures. These rather large values for the pion damping are taken into account  with $\gamma_2(T)$ in our second model (cf.~Eq.~\eqref{eq:gamT2}). As written in the main text, we choose the two parameters $\alpha$ and $T_{**}$ to reproduce the qualitative shape of the reconstructed damping factor of exponents of Ref.~\cite{Lowdon:2022xcl}, which leads to the values $\alpha \approx 5.5$ and $T_{**} \approx 200$~MeV.

There are, however, important caveats in this naive identification. In the spectral reconstruction of Ref.~\cite{Lowdon:2022xcl} a form for the spectral function is used that is different from our Breit-Wigner form of the present work. (The Breit-Wigner form originates from the Ohmic spectral distribution of the heat bath.) This means that, strictly speaking, the damping factor exponents of Ref.~\cite{Lowdon:2022xcl} can not be directly identified with our damping constant $\gamma(T)$. Nevertheless, we consider it to be important to study the situation where the pion damping is as large as the data from Ref.~\cite{Lowdon:2022xcl} suggests. Hence, for the scope of this work, we assume that at least the order of magnitude of our pion damping roughly matches to the damping-factor exponents of Ref.~\cite{Lowdon:2022xcl}. For the future it would be interesting to incorporate the thermoparticle assumption into the quark-meson model consistently, and study its effects both on the static equilibrium properties and the dynamical properties. 

\bibliographystyle{h-physrev3}
\bibliography{biblio.bib}

\begin{thebibliography}{10}

\bibitem{Fischer:2018sdj}
C.~S. Fischer,
\newblock Prog. Part. Nucl. Phys. {\bf 105}, 1 (2019), 1810.12938.

\bibitem{Fu:2019hdw}
W.-j. Fu, J.~M. Pawlowski, and F.~Rennecke,
\newblock Phys. Rev. D {\bf 101}, 054032 (2020), 1909.02991.

\bibitem{Gao:2020qsj}
F.~Gao and J.~M. Pawlowski,
\newblock Phys. Rev. D {\bf 102}, 034027 (2020), 2002.07500.

\bibitem{Gunkel:2021oya}
P.~J. Gunkel and C.~S. Fischer,
\newblock Phys. Rev. D {\bf 104}, 054022 (2021), 2106.08356.

\bibitem{Braun:2022jme}
J.~Braun, A.~Gei\ss{}el, and B.~Schallmo,
\newblock SciPost Phys. Core {\bf 7}, 015 (2024), 2206.06328.

\bibitem{Tripolt:2013jra}
R.-A. Tripolt, N.~Strodthoff, L.~von Smekal, and J.~Wambach,
\newblock Phys. Rev. D {\bf 89}, 034010 (2014), 1311.0630.

\bibitem{Tripolt:2014wra}
R.-A. Tripolt, L.~von Smekal, and J.~Wambach,
\newblock Phys. Rev. D {\bf 90}, 074031 (2014), 1408.3512.

\bibitem{Jung:2016yxl}
C.~Jung, F.~Rennecke, R.-A. Tripolt, L.~von Smekal, and J.~Wambach,
\newblock Phys. Rev. D {\bf 95}, 036020 (2017), 1610.08754.

\bibitem{Tripolt:2018qvi}
R.-A. Tripolt, J.~Weyrich, L.~von Smekal, and J.~Wambach,
\newblock Phys. Rev. D {\bf 98}, 094002 (2018), 1807.11708.

\bibitem{Tripolt:2020irx}
R.-A. Tripolt, D.~H. Rischke, L.~von Smekal, and J.~Wambach,
\newblock Phys. Rev. D {\bf 101}, 094010 (2020), 2003.11871.

\bibitem{Tripolt:2021jtp}
R.-A. Tripolt, C.~Jung, L.~von Smekal, and J.~Wambach,
\newblock Phys. Rev. D {\bf 104}, 054005 (2021), 2105.00861.

\bibitem{Pawlowski:2017gxj}
J.~M. Pawlowski, N.~Strodthoff, and N.~Wink,
\newblock Phys. Rev. D {\bf 98}, 074008 (2018), 1711.07444.

\bibitem{Horak:2020eng}
J.~Horak, J.~M. Pawlowski, and N.~Wink,
\newblock Phys. Rev. D {\bf 102}, 125016 (2020), 2006.09778.

\bibitem{Horak:2022aza}
J.~Horak, J.~M. Pawlowski, and N.~Wink,
\newblock SciPost Phys. {\bf 15}, 149 (2023), 2210.07597.

\bibitem{Braun:2022mgx}
J.~Braun {\em et~al.},
\newblock SciPost Phys. Core {\bf 6}, 061 (2023), 2206.10232.

\bibitem{Horak:2022myj}
J.~Horak, J.~M. Pawlowski, and N.~Wink,
\newblock (2022), 2202.09333.

\bibitem{Horak:2023hkp}
J.~Horak, F.~Ihssen, J.~M. Pawlowski, J.~Wessely, and N.~Wink,
\newblock (2023), 2303.16719.

\bibitem{Horak:2023xfb}
J.~Horak {\em et~al.},
\newblock Phys. Rev. D {\bf 107}, 076019 (2023), 2301.07785.

\bibitem{Eichmann:2023tjk}
G.~Eichmann {\em et~al.},
\newblock Phys. Rev. D {\bf 109}, 096024 (2024), 2310.16353.

\bibitem{Berges:2012ty}
J.~Berges and D.~Mesterhazy,
\newblock Nucl. Phys. B Proc. Suppl. {\bf 228}, 37 (2012), 1204.1489.

\bibitem{Huelsmann:2020xcy}
S.~Huelsmann, S.~Schlichting, and P.~Scior,
\newblock Phys. Rev. D {\bf 102}, 096004 (2020), 2009.04194.

\bibitem{Roth:2021nrd}
J.~V. Roth, D.~Schweitzer, L.~J. Sieke, and L.~von Smekal,
\newblock Phys. Rev. D {\bf 105}, 116017 (2022), 2112.12568.

\bibitem{Canet:2006xu}
L.~Canet and H.~Chate,
\newblock J. Phys. {\bf 40}, 1937 (2007), cond-mat/0610468.

\bibitem{Canet:2011wf}
L.~Canet, H.~Chate, and B.~Delamotte,
\newblock J. Phys. A {\bf 44}, 495001 (2011), 1106.4129.

\bibitem{Mesterhazy:2013naa}
D.~Mesterh\'azy, J.~H. Stockemer, L.~F. Palhares, and J.~Berges,
\newblock Phys. Rev. B {\bf 88}, 174301 (2013), 1307.1700.

\bibitem{Mesterhazy:2015uja}
D.~Mesterh\'azy, J.~H. Stockemer, and Y.~Tanizaki,
\newblock Phys. Rev. D {\bf 92}, 076001 (2015), 1504.07268.

\bibitem{Duclut:2016jct}
C.~Duclut and B.~Delamotte,
\newblock Phys. Rev. E {\bf 95}, 012107 (2017), 1611.07301.

\bibitem{Tan:2021zid}
Y.-y. Tan, Y.-r. Chen, and W.-j. Fu,
\newblock SciPost Phys. {\bf 12}, 026 (2022), 2107.06482.

\bibitem{Roth:2023wbp}
J.~V. Roth and L.~von Smekal,
\newblock JHEP {\bf 10}, 065 (2023), 2303.11817.

\bibitem{Chen:2023tqc}
Y.-r. Chen, Y.-y. Tan, and W.-j. Fu,
\newblock Phys. Rev. D {\bf 109}, 094044 (2024), 2312.05870.

\bibitem{Batini:2023nan}
L.~Batini, E.~Grossi, and N.~Wink,
\newblock Phys. Rev. D {\bf 108}, 125021 (2023), 2309.06586.

\bibitem{Tan:2024fuq}
Y.-y. Tan, Y.-r. Chen, W.-j. Fu, and W.-J. Li,
\newblock (2024), 2403.03503.

\bibitem{Roth:2024rbi}
J.~V. Roth, Y.~Ye, S.~Schlichting, and L.~von Smekal,
\newblock JHEP {\bf 01}, 118 (2025), 2403.04573.

\bibitem{Chen:2024lzz}
Y.-r. Chen, Y.-y. Tan, and W.-j. Fu,
\newblock (2024), 2406.00679.

\bibitem{Roth:2024hcu}
J.~V. Roth, Y.~Ye, S.~Schlichting, and L.~von Smekal,
\newblock (2024), 2409.14470.

\bibitem{Hohenberg:1977ym}
P.~C. Hohenberg and B.~I. Halperin,
\newblock Rev. Mod. Phys. {\bf 49}, 435 (1977).

\bibitem{Shen:2020jya}
L.~Shen, J.~Berges, J.~M. Pawlowski, and A.~Rothkopf,
\newblock Phys. Rev. D {\bf 102}, 016012 (2020), 2003.03270.

\bibitem{Meistrenko:2020nwx}
A.~Meistrenko, H.~van Hees, and C.~Greiner,
\newblock Annals Phys. {\bf 431}, 168555 (2021), 2007.09929.

\bibitem{Sieberer:2015hba}
L.~M. Sieberer, A.~Chiocchetta, A.~Gambassi, U.~C. T\"auber, and S.~Diehl,
\newblock Phys. Rev. B {\bf 92}, 134307 (2015), 1505.00912.

\bibitem{Altland:2020lbb}
A.~Altland, M.~Fleischhauer, and S.~Diehl,
\newblock Phys. Rev. X {\bf 11}, 021037 (2021), 2007.10448.

\bibitem{Rajagopal:1992qz}
K.~Rajagopal and F.~Wilczek,
\newblock Nucl. Phys. B {\bf 399}, 395 (1993), hep-ph/9210253.

\bibitem{Son:2004iv}
D.~T. Son and M.~A. Stephanov,
\newblock Phys. Rev. D {\bf 70}, 056001 (2004), hep-ph/0401052.

\bibitem{CALDEIRA1983587}
A.~Caldeira and A.~Leggett,
\newblock Physica A: Statistical Mechanics and its Applications {\bf 121}, 587
  (1983).

\bibitem{kamenev_2011}
A.~Kamenev,
\newblock {\em Field Theory of Non-Equilibrium Systems} (Cambridge University
  Press, 2011).

\bibitem{Das:1997gg}
A.~K. Das,
\newblock {\em {Finite Temperature Field Theory}} (World Scientific, New York,
  1997).

\bibitem{Gies:2006wv}
H.~Gies,
\newblock Lect. Notes Phys. {\bf 852}, 287 (2012), hep-ph/0611146.

\bibitem{bjorken1965relativistic}
J.~Bjorken and S.~Drell,
\newblock {\em Relativistic Quantum Fields} (McGraw-Hill, 1965).

\bibitem{schwabl2008advanced}
F.~Schwabl, R.~Hilton, and A.~Lahee,
\newblock {\em Advanced Quantum Mechanics} (Springer Berlin Heidelberg, 2008).

\bibitem{Berges:1997eu}
J.~Berges, D.~U. Jungnickel, and C.~Wetterich,
\newblock Phys. Rev. D {\bf 59}, 034010 (1999), hep-ph/9705474.

\bibitem{Schaefer:2004en}
B.-J. Schaefer and J.~Wambach,
\newblock Nucl. Phys. A {\bf 757}, 479 (2005), nucl-th/0403039.

\bibitem{Strodthoff:2011tz}
N.~Strodthoff, B.-J. Schaefer, and L.~von Smekal,
\newblock Phys. Rev. D {\bf 85}, 074007 (2012), 1112.5401.

\bibitem{Tripolt:2017zgc}
R.-A. Tripolt, B.-J. Schaefer, L.~von Smekal, and J.~Wambach,
\newblock Phys. Rev. D {\bf 97}, 034022 (2018), 1709.05991.

\bibitem{Resch:2017vjs}
S.~Resch, F.~Rennecke, and B.-J. Schaefer,
\newblock Phys. Rev. D {\bf 99}, 076005 (2019), 1712.07961.

\bibitem{Grossi:2021ksl}
E.~Grossi, F.~J. Ihssen, J.~M. Pawlowski, and N.~Wink,
\newblock Phys. Rev. D {\bf 104}, 016028 (2021), 2102.01602.

\bibitem{Otto:2022jzl}
K.~Otto, C.~Busch, and B.-J. Schaefer,
\newblock Phys. Rev. D {\bf 106}, 094018 (2022), 2206.13067.

\bibitem{Ihssen:2023xlp}
F.~Ihssen, J.~M. Pawlowski, F.~R. Sattler, and N.~Wink,
\newblock (2023), 2309.07335.

\bibitem{Wetterich:1992yh}
C.~Wetterich,
\newblock Phys. Lett. B {\bf 301}, 90 (1993), 1710.05815.

\bibitem{Litim:2001up}
D.~F. Litim,
\newblock Phys. Rev. D {\bf 64}, 105007 (2001), hep-th/0103195.

\bibitem{Grossi:2019urj}
E.~Grossi and N.~Wink,
\newblock SciPost Phys. Core {\bf 6}, 071 (2023), 1903.09503.

\bibitem{Ihssen:2023qaq}
F.~Ihssen, F.~R. Sattler, and N.~Wink,
\newblock Phys. Rev. D {\bf 107}, 114009 (2023), 2302.04736.

\bibitem{Pelaez:2015qba}
J.~R. Pelaez,
\newblock Phys. Rept. {\bf 658}, 1 (2016), 1510.00653.

\bibitem{Goity:1989gs}
J.~L. Goity and H.~Leutwyler,
\newblock Phys. Lett. B {\bf 228}, 517 (1989).

\bibitem{Leutwyler:1990uq}
H.~Leutwyler and A.~V. Smilga,
\newblock Nucl. Phys. B {\bf 342}, 302 (1990).

\bibitem{Lowdon:2022xcl}
P.~Lowdon and O.~Philipsen,
\newblock JHEP {\bf 10}, 161 (2022), 2207.14718.

\bibitem{Braun:2018svj}
J.~Braun, M.~Leonhardt, and J.~M. Pawlowski,
\newblock SciPost Phys. {\bf 6}, 056 (2019), 1806.04432.

\bibitem{Haensch:2023sig}
M.~Haensch, F.~Rennecke, and L.~von Smekal,
\newblock (2023), 2308.16244.

\end{thebibliography}
\end{document}